\documentclass[twocolumn,10pt]{svjour3}

\usepackage{mathptmx}
\usepackage{mathrsfs}	%fancy curly font
\usepackage{bm}		% boldfaced math
\usepackage{amsmath}    % need for subequations
\usepackage{graphicx}   % need for figures
\usepackage{verbatim}   % useful for program listings
\usepackage{color}      % use if color is used in text
\usepackage{subfigure}  % use for side-by-side figures
\usepackage{hyperref}   % use for hypertext links, including those to external documents and URLs
\usepackage{graphics,graphicx}
\usepackage{epstopdf}
\usepackage{times}
\usepackage{amssymb}
\usepackage{amsfonts}
\usepackage{amsbsy}
\usepackage{multirow}
\usepackage{array,tabularx}
\usepackage{color}
\usepackage{float}

\usepackage[round, sort, comma, authoryear]{natbib}

\graphicspath{{./images/}}

\newcommand{\SF}{\bfseries \sffamily}

\begin{document}

%\noindent Notes to self:
%
%
%\noindent Notes for Next Meeting:
%
%-  Permission to use high school network?
%
%-  Figures to top or bottom of page

\title{A Modular Multiscale Approach to Overlapping Community Detection}
\author{M. Brutz \and F. G. Meyer}
\institute{M. Brutz \at Department of Applied Mathematics, University of Colorado, Boulder, CO, USA \\\email{michael.brutz@colorado.edu} \and F. G. Meyer \at Department of Electrical Engineering, University of Colorado, Boulder, CO, USA  \\\email{fmeyer@colorado.edu}}
\maketitle

%=================================================================
%   ABSTRACT
%=================================================================
\begin{abstract}

In this work we address the problem of detecting overlapping communities in social networks. Because the word "community" is an ambiguous term, it is necessary to quantify what it means to be a community within the context of a particular type of problem.  Our interpretation is that this quantification must be done at a minimum of three scales.  These scales are at the level of:  individual nodes, individual communities, and the network as a whole.  Each of these scales involves quantitative features of community structure that are not accurately represented at the other scales, but are important for defining a particular notion of community.  Our work focuses on providing sensible ways to quantify what is desired at each of these scales for a  notion of community applicable to social networks, and using these models to develop a community detection algorithm.  Appealing features of our approach is that it naturally allows for nodes to belong to multiple communities, and is computationally efficient for large networks with low overall edge density.  The scaling of the algorithm is $O(N~\overline{k^2} + \overline{N_{com}^2})$, where $N$ is the number of nodes in the network, $\overline{N_{com}^2}$ is the average squared community size, and $\overline{k^2}$ is the expected value of a node's degree squared.  Although our work focuses on developing a computationally efficient algorithm for overlapping community detection in the context of social networks, our primary contribution is developing a methodology that is highly modular and can easily be adapted to target specific notions of community.


\end{abstract}

%Two primary perspectives are at the scale of individual nodes, and at the scale of individual communities, with a secondary perspective at the scale of the network as a whole.  

%=================================================================
%   INTRODUCTION
%=================================================================
\section{Introduction}

A number of real-world systems are mathematically represented by graphs, where nodes represent agents in the system and edges represent connections between agents.  A common feature of interest in such systems comes from finding groups of nodes that can be considered as communities within the network.  Although community detection is often referred to as though it is a single problem, a more accurate description is that it is a body of related problems, owing to the fact that the notion of what it means to be a "community" involves different concepts in different contexts.  

When adopting the view that communities are not simply defined by nodes but the connections between nodes, a sensible approach to detecting communities is to cluster the edges present in the network.  The first paper to explicitly acknowledge this point of view was \citep{EvLa09}, where communities are formed by partitioning the line graph of the network.  However, edge clustering methods predate this by several years.  Although it was not viewed in terms of clustering edges at the time of its publication, clique percolation \citep{Pal05} is perhaps one of the earliest examples of such methods, as cliques can be equivalently described as sets of completely interconnected nodes or edges.  

In the clique percolation method, each node is described by the cliques it is a member of, and these cliques serve as the "atoms" from which to build community "molecules".  The cliques provide a mechanism for representing a localized feature of a community structure.  These localized features are then agglomerated to carve out communities by what the authors of \citep{Pal05} call a "percolation" process.  This process consists of repeatedly adding all other cliques to the community that differ by (at most) one node from a clique already present in the community.

A natural generalization of the clique percolation methodology is to derive a node's local affinity for one, or multiple communities from its local perspective of the network itself.  In such an approach, it is useful to represent the local neighborhood of each node by an "egonet".  An \textit{egonet} is the network restricted to just the set of nodes a given node is connected to, along with all of the edges between the nodes in that set, where the node this net is built around is called the \textit{egocentric node}. 

This generalization serves as the motivation for the collective friendship group inference method \citep{FriendshipGroup10}.  With this method, instead of describing each node by the cliques it is a member of, each node is described by what are called friendship circles.  A \textit{friendship circle} is any set of edges in an egonet that satisfy the definition of community on the network as a whole when the egocentric node is removed.  These friendship circles then play the role of cliques in the aforementioned method, and communities are formed via an equivalent percolation process.  

Although our methodology is most closely related to the preceding edge clustering methods, we also incorporate elements based on quality function optimization and spectral clustering \citep{SpectralClusteringOverview, SpectralClusteringComparativeAnalysis}.  The interested reader can find a comprehensive overview of these approaches (and many others) in \citep{Fortunato10}, and an analysis of overlapping community detection methods in \citep{Xi13}.

%=================================================================
%   My Approach
%=================================================================
\section{Proposed Approach}

\begin{figure}[h!] 
 \begin{center}
	\includegraphics [width=0.45\textwidth] {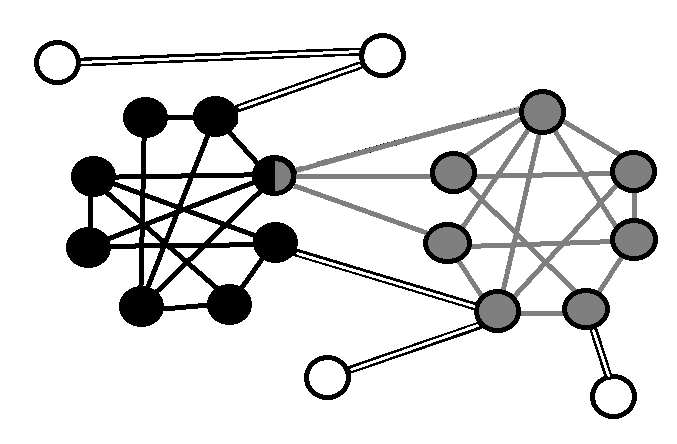} 
  \end{center}
   \caption{ Example network with community structure.}
\label{fig:AlgorithmExplainedBaseNetwork}
\end{figure}

Our work focuses on community detection with respect to social networks representable by undirected / unweighted graphs.  We assume communities are fundamentally defined in terms of the properties of the edges comprising it, so we approach community detection as an edge clustering problem.  As clustering edges naturally allows a node to belong to multiple communities based on how the edges it is connected to are clustered, this conforms well with a common feature of social networks where communities can overlap with one another. 

Using the graph depicted in Figure \ref{fig:AlgorithmExplainedBaseNetwork} as a model network, several questions naturally arise on how to describe the community structure present.  Firstly, what differentiates edges that connect a node to members of the same community from those that connect them to non-community members; what properties does a community possess at the scale of individual nodes?  Secondly, what differentiates the sets of black and gray edges from the white ones; what properties does a community possess at the scale of individual communities?  Lastly, what should be done with nodes that do not decisively belong to any particular community; what properties does the set of all communities possess at the scale of the entire network?  

These questions confront the multiple scales that intrinsically define communities: the scale of individual nodes, the scale of individual communities, and the scale of the entire network.  In this work, we attempt to provide logical answers to these foundational questions in the context of identifying overlapping communities in social networks.  Because our approach is highly modular, one can easily modify the specific quantitative model for community structure at each scale as appropriate for a given community detection problem.  This opens the door for creating community detection algorithms capable of searching for targeted notions of community that respect the context of the problem.

Let $N$ be the number of nodes in the network, $\overline{N_{com}^2}$ be the average squared community size, and $\overline{k^2}$ be the expected value of a node's degree squared.  An overview of our algorithm and the computational cost of each step is presented in Figure \ref{alg:CD}.

\begin{figure}[h]
\makebox{\SF Community Detection Algorithm} \\
\rule[0mm]{\linewidth}{0.5pt} 
  \begin{itemize}
  \item [] {\SF Input}: 
    \begin{itemize}
    	\item Adjacency matrix for the network
    \end{itemize}
  \item []{\SF Algorithm:}
    \begin{enumerate}
    	\item  Detect the sets of edges that will be used to describe each node in the network as described in Section \ref{sec:NodeLevelPerspective}.  
	\begin{itemize}
		\item  Cost:  $O(N~\overline{k^2})$
	\end{itemize}
	\item  Cluster the edge sets to form communities as described in Section \ref{sec:CommunityLevelPerspective}. 
	\begin{itemize}
		\item  Cost:  $O(\overline{N_{com}^2})$
	\end{itemize} 
	\item  If any nodes in the network remain unclustered, attach them to the community they share the most connections with. 
	\begin{itemize}
		\item  Cost:  Negligible
	\end{itemize}  
	\item  Prune out the smaller communities detected subject to the constraint that all nodes are represented in at least one community, as described in Section \ref{sec:NetworkLevelPerspective}.  
	\begin{itemize}
		\item  Cost:  Negligible
	\end{itemize}  
    \end{enumerate}
  \item  [] {\SF Output:}  Sets of nodes comprising communities detected on network.
  \end{itemize}
  \rule[0mm]{\linewidth}{0.5pt} 
  \caption{Community detection algorithm
    \label{alg:CD}}
\end{figure}

%\begin{itemize}
%\item
%\end{itemize}

%

%The computational cost of our algorithm is dominated by steps 1 and 2, which are $O(N~\overline{k^2})$ and $O(\overline{N_{com}^2})$ respectively.  The cost of step 1 is incurred in calculating subsets of eigenvectors for each of the adjacency matrices corresponding to each node's egonet.  The cost of step 2 is primarily driven by searching for appropriate edge sets to cluster with the communities being formed.  

%[[[  NOTE:  I AM NOT SEEING WHERE WOULD BE A GOOD PLACE TO DISCUSS COMPUTATIONAL COSTS IN THIS SECTION ]]]

\subsection{Node Scale Features of Community Structure:  Edge Descriptor Sets}
\label{sec:NodeLevelPerspective}

\begin{figure}[t] 
 \begin{center}
	\includegraphics [width=0.45\textwidth] {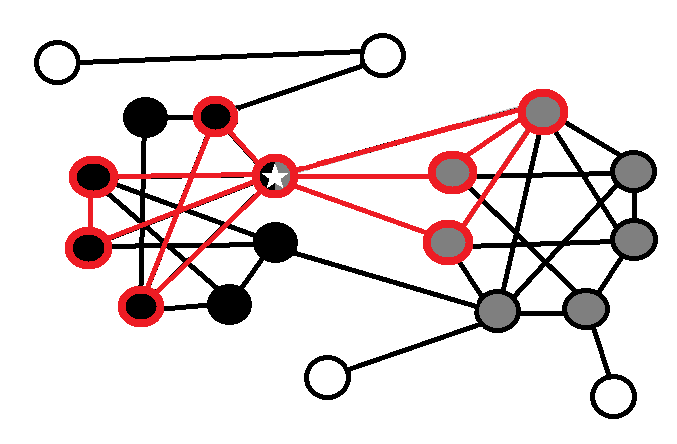} 
  \end{center}
   \caption{  A depiction of the egonet for the starred node in the network; all of the red nodes/edges are included in the egonet.}
\label{fig:AlgorithmExplainedNodeLevel}
\end{figure}

Because we are interested in identifying communities from the arrangement of edges, we introduce the notion of \textit{edge descriptor sets} as a general term for using sets of edges to describe community structure local to each node.  The cliques (or friendship circles) a node belongs to, would be specific examples such edge descriptor sets.

In our approach, we assume that a node is more likely to belong to a community if it has many mutual friends within that community. This suggests that edges linking members of the same community should be densely inter-connected. The prototypical example of such a cluster of edges would be a clique. Furthermore, we note that if a node belongs to several relatively large cliques that are mutually disjoint, then this node may be at the intersection of multiple communities.  Guided by these simple heuristic principles, we propose to extract sets of edges that are both densely connected and largely disjoint from each other from each node's egonet to serve as the edge descriptor sets for that node.  Each set can be thought of as encoding the fine-scale features of community structures present in the network.

%assume a node's localized perspective of a community it belongs to is reflected by having multiple mutual friends in that community.  This suggests that sets of edges modeling these perspectives should be densely interconnected, as with a clique.  We now note that when a node belongs to relatively large cliques that are disjoint from one another, this can serve as an indication that the node is a member of multiple communities.  Using this intuition as a guide, we aim to extract sets of edges that are both densely connected and largely disjoint from each other from each node's egonet. Each set of edges comprises an edge descriptor set that encodes the fine-scale connectivity around each node in the network. 
 
We approach the task of extracting edge descriptor sets by first using a simple spectral clustering method to sparsify the local egonet defined around each node. The goal of the sparsification is to reveal the largest disjoint cliques that are present in the egonet. Finally, a more advanced spectral clustering method is used to construct the edge descriptor sets associated with the node in question.  We begin our discussion with the latter of these two processes, because it is more involved and provides a natural introduction to the former.

%The sparsification process is accomplished by finding good candidates for the largest cliques each node in an egonet belongs to, and removing any connections that involve the node but are not components of that largest clique from the egonet.  Once this matrix has been fully sparsified, we use a separate spectral clustering method to find disjoint sets of edges that approximately form cliques involving the egocentric node.  We begin our discussion with the latter of these two processes, because it is more involved and provides a natural introduction to the former.

\subsubsection{ICM-Matrices}
\label{sec:ICM}

In this section, we examine the spectral properties of idealized egonets formed by cliques that are only connected through a single (egocentric) vertex. While this situation may appear unrealistic, we explain in the next section how to extract such subgraphs from the original graph. Our present goal is simply to identify and extract the corresponding cliques; we propose a spectral approach to solve this problem.  

Let us consider the sub-matrix of the network adjacency matrix that describes the local egonet. A trivial re-indexing of the nodes allows us to represent this submatrix as a block-diagonal matrix, where each block is a clique. The blocks do not overlap, but there is a row (and corresponding column) of ones to describe the connection of the egocentric node to all the cliques. We can assume that the row and corresponding column associated with the egocentric node are the first row and column, respectively. Instead of working directly with this matrix, we propose to make some slight modifications that will boost the spectral approach.  The modifications are as follows: we add self connections to all the nodes if not already present, and we scale all connections to the egocentric node by a small parameter, $\delta$.  We will call these modified matrices \textit{ideal community member matrices} for the sake of discussion, or \textit{ICM-matrices} for short.  

\begin{figure}[h!] 
 \begin{center}
	\includegraphics [width=0.45\textwidth] {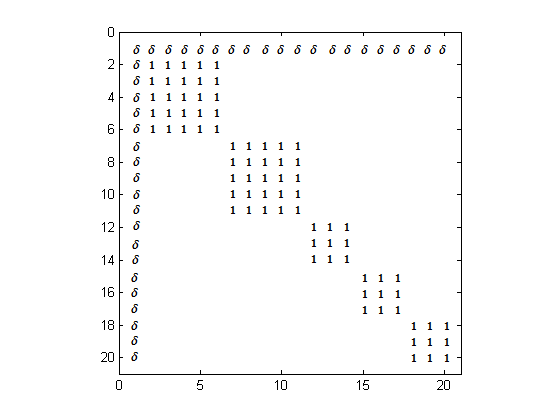} 
  \end{center}
   \caption{The non-zero values of an ICM-matrix for a node belonging to two cliques with 6 members, and three cliques of four members.  }
\label{fig:IdealCMmatrix}
\end{figure}

Let $A$ be an ICM-matrix with $m$ blocks (each corresponding to a clique) along its diagonal, where the size of the $i^{th}$ block is $k_i \times k_i$.  Without loss of generality, we assume the indices are arranged such that $k_j \le k_i$ whenever $j>i$ so that larger indices correspond to smaller blocks.  Also assume that the first index corresponds to the egocentric node.  We will represent the set of indices for the $k^{th}$ block by $V_k$.  With this notation in place, $A$ is defined by Equation \eqref{eq:ADef}, and an example of the structure of such a matrix is given in Figure \ref{fig:IdealCMmatrix}.

\begin{equation}
	A(i,j) = 
	\begin{cases}
		 & \delta, ~~~~~  ~~~ \text{if}~ i=1~\text{or}~j=1 , \\
		 & 1, ~~~~~   ~~~ \forall i,j \in V_k , ~~  k=1,...,m  ,   \\
	            & 0, ~~~~~ ~~~ \text{otherwise}  .
	\end{cases}
	\label{eq:ADef}
\end{equation}

%In order to identify the blocks associated with the cliques, we compute the eigenvectors of $A$ that are associated with positive eigenvalues. 

Our focus will be on the eigenvectors of these matrices with corresponding positive eigenvalues, as they will be the ones we will use for identifying the cliques via spectral clustering.  Let $P = \{i | \lambda_i > 0\}$ be the positive eigenvalues, and let $\{\mathbf{x}_p |  p \in P\}$ be the associated eigenvectors.  Assume that the eigenvalues are ordered so that $\lambda_j \le \lambda_i$ if $i>j$ (e.g.  $\mathbf{x}_1$ is the dominant eigenvector).  Let $\mathbf{x}_p(1)$ be the value of the first entry in such an eigenvector.  Using a simple symmetry argument, one can show that each $\mathbf{x_p}$ is constant over each clique $i$; $\mathbf{x_p}(l \in V_i) = r_i$, for each vertex, $l$, in the set of vertices, $V_i$, in block $i$. 

We now examine the properties of the entries of $\mathbf{x_p}$ by explicitly looking at the system of equations resulting from the constraint $\lambda_p ~ \mathbf{x}_p = A~\mathbf{x}_p $.  For any $j \in V_i$, this constraint takes the following form,

%\begin{subequations}
%	\begin{align}
%		\lambda_p~\mathbf{x}_p(1) &= \delta ~ \sum_{i} \mathbf{x}_p(i) \label{eq:FirstRowa} \\
%		~ &= \delta ~ \mathbf{x}_p(1)+  \delta ~ \sum_{i=1}^{m} ~ \sum_{j=1}^{k_i } ~ r_i  \label{eq:FirstRowb} \\
%	          ~  &= \delta ~ \mathbf{x}_p(1) + \delta ~ \sum_{i=1}^{m} k_i ~ r_i  \label{eq:G1}  
%	\end{align}
%	\label{eq:FirstRow}
%\end{subequations}

\begin{equation}
		\lambda_p~\mathbf{x}_p(j \in V_i) = \delta ~ \mathbf{x}_p(1) + \sum_{l ~ \in ~ V_i} \mathbf{x}_p(l), \label{eq:OtherRowa}
\end{equation}

\noindent and since $\mathbf{x_p}$ is constant and equal to $r_i$ over each clique, we obtain

\begin{equation}
		\lambda_p~r_i = \delta ~ \mathbf{x}_p(1) + k_i ~ r_i, 
\end{equation}

\noindent or

\begin{equation}
		(\lambda_p - k_i) r_i = \delta ~ \mathbf{x}_p(1) .   \label{eq:G2}  
\end{equation}

As Equation \eqref{eq:G2} holds for any row indices, $i$ and $j$, this leads us to Equation \eqref{eq:G3},

\begin{equation}
	r_i = \frac{\lambda_p - k_j }{\lambda_p - k_i } r_j .
\label{eq:G3}
\end{equation}

The importance of Equation \eqref{eq:G3} is that it shows that $|r_i| > |r_j|$ if $|\lambda_p - k_i| < |\lambda_p - k_j| $.  In other words, the closer the number of members in the block corresponding to the $i^{th}$ clique is to $\lambda_p$, the larger the magnitude of $r_i$ in $\mathbf{x}_p$ and the easier it will be to pull out members of distinct cliques via spectral clustering. Now note that by setting $\delta$ to a small value, the ICM-matrix is a small perturbation of a block diagonal matrix.  As the positive eigenvalues of a block diagonal matrix correspond directly to the sizes of the blocks present within the matrix, this will cause the eigenvalues of the ICM-matrix to be small perturbations of $k_i$.  This implies that for each block of size $k_i$ there exists an eigenvalue, $\lambda$, such that $\lambda \approx k_i$.  For the results presented in this paper, we use $\delta = 1/N_{ego}$, where $N_{ego}$ is the number of nodes in the egonet.

\begin{figure}[t] 
 \begin{center}
	\includegraphics [width=0.5\textwidth] {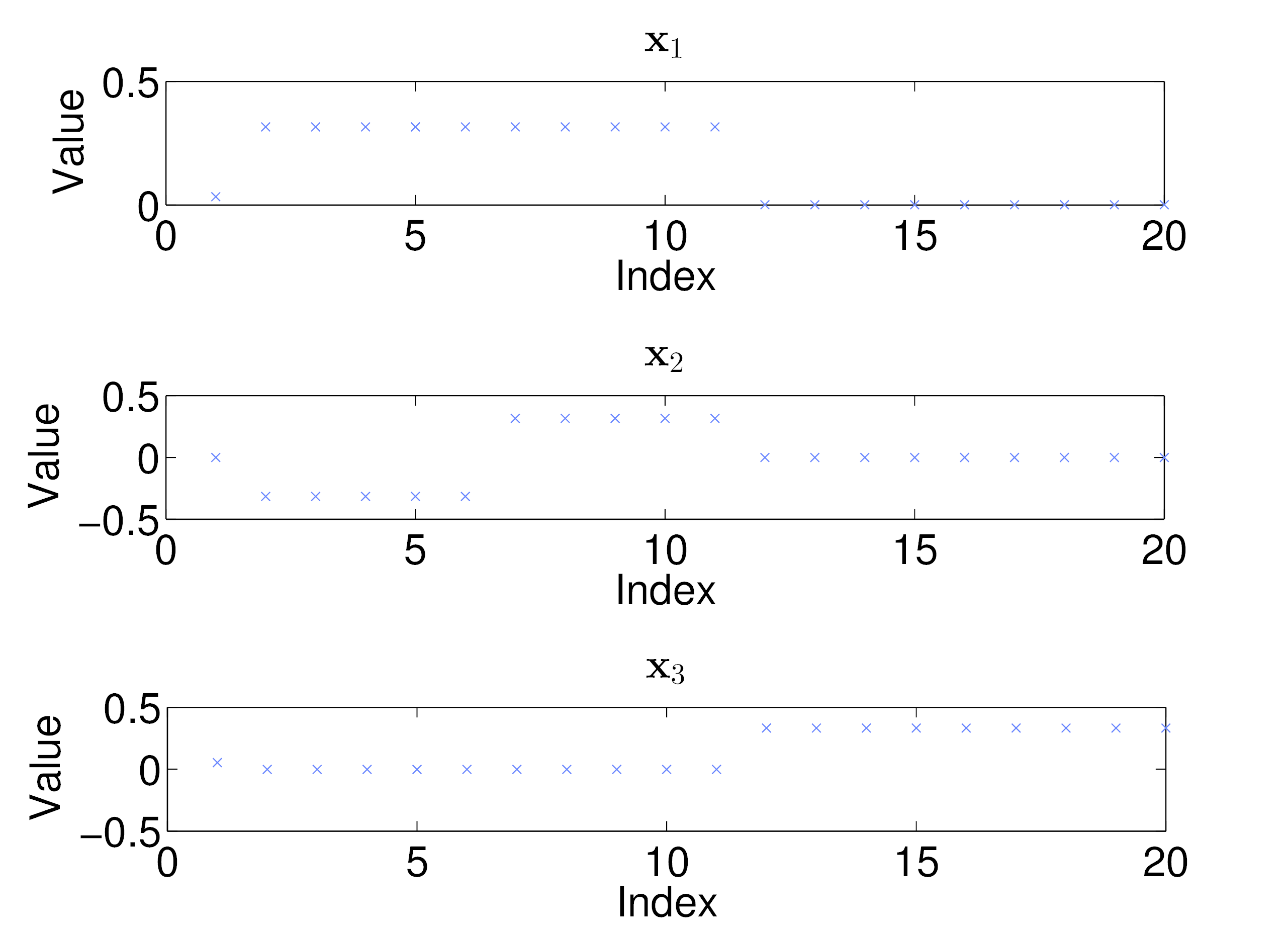} 
  \end{center}
   \caption{First three dominant eigenvectors for the ICM-matrix depicted in Figure \ref{fig:IdealCMmatrix}, where $\delta = 1/N_{ego}$ and $N_{ego}$ is the number of nodes involved in the egonet.  }
\label{fig:IdealCMmatrixEigVecs}
\end{figure}

For illustrative purposes, the first three dominant eigenvectors of the matrix depicted in Figure \ref{fig:IdealCMmatrix} are shown in Figure \ref{fig:IdealCMmatrixEigVecs}.  Note that the components of the eigenvectors clearly reveal each clique.  However, this property does not extend to eigenvectors without positive eigenvalues.  Indeed, the nullspace vectors are quite noisy and are detrimental to use for spectral clustering of the egonet.

For the case of $\lambda_1$, we can also prove that $r_i \ge r_j$ when $j>i$.  This follows from the fact that all of the entries of $A$ are non-zero and the power method will converge to the dominant eigenvector.  As the power method will converge regardless of the initial starting vector, we can take a vector with all non-negative entries as our initial guess and the power method iteration will preserve this property.  This implies that the magnitude of $\mathbf{x}_1(i)$  increases with the size of the clique in which the vertex $i$ belongs to.  This is important to the egonet sparsification algorithm we discuss in the next section.

We are now in a position to define the edge descriptor sets associated with each node in our approach. Each edge descriptor set is comprised of densely connected subnetworks of the egonet. Formally, we define the densely connected subnetworks as follows.  We begin by extracting the node's egonet from the network, and form its corresponding adjacency matrix.  Using the methods described in the next section, we sparsify this adjacency matrix so that the remaining connections closely resemble the structure of an ICM-matrix.  Next, we compute the eigenvectors associated with the largest positive eigenvalues of the sparsified matrix.  We then use these eigenvectors to embed the egonet into a metric space, treating the value of each eigenvector at a given index as providing a spacial ordinate for the corresponding vertex \citep{Br03}.  With each vertex having spacial coordinates, we then apply k-means clustering in order to find clusters of vertices.  As each vertex is also connected to the egocentric node, each cluster of vertices represents a cluster of edges to use as a potential edge descriptor set.  Finally, before accepting a cluster as an edge descriptor set, we additionally check that the set of nodes involved has 90\% or greater edge density between them to ensure they approximately form a clique.  Although these edge descriptor sets will often be referred to as "cliques" for simplicity throughout this paper, we only require them to be very densely connected rather than fully connected.

To determine both the number of eigenvectors to use for the spectral embedding and the number of clusters, we use the following set of heuristics.  Presumably, the largest cliques a node belongs to are the most important ones to accurately capture for describing that node.  As larger cliques correspond to larger eigenvalues in an ICM-matrix, we estimate the number of coordinates for the spectral embedding and the number of clusters as the number of eigenvalues that are greater than one tenth of the largest eigenvalue.  This allows us to recover all the largest cliques an egocentric node belongs to, while guarding against involving near nullspace vectors for the clustering process.

\subsubsection{Matrix Sparsification Algorithm}

\begin{figure}[h!]
  \begin{center}
    \subfigure [ Egonet for node A] { \label{fig:SparsificationAlgorithmExplained} \includegraphics [width=0.23\textwidth] {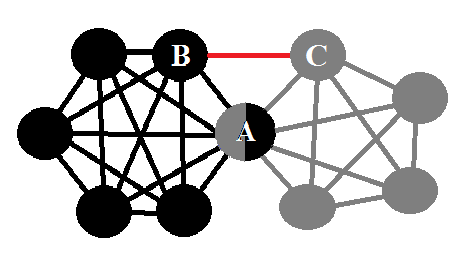} }%
    \subfigure [Sub-egonet of A for node B] { \label{fig:SparsificationAlgorithmExplained_SubEgo} \includegraphics [width=0.23\textwidth] {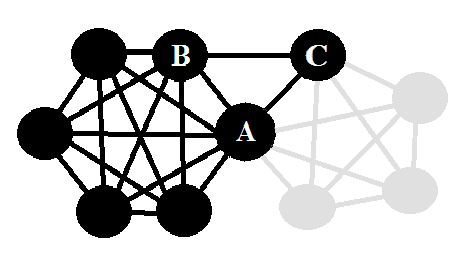} }%
  \end{center}
   \caption{  a) The egonet of the node A is formed by the union of the gray and black nodes and corresponding edges. However, the cliques lying within each community are not completely disjoint, as the red edge connects node B in the black community to node C in the gray community.   b)  An example of a sub-egonet, where the black nodes and edges represent the sub-egonet of A for sub-egocentric node B.}
\end{figure}

In practice, it is unrealistic to expect to encounter egonets corresponding directly to the ideal community member matrices described in the previous section.  A more realistic assumption is that the cliques falling in different communities have some random edges connecting them, as in Figure \ref{fig:SparsificationAlgorithmExplained}.  However, adding such random connections to ICM-matrices can drastically alter their eigenspace properties, so this problem must be addressed.  

To this end, we develop a method capable of removing such connections with high accuracy.  This is accomplished by considering each node present in the egonet, and removing any edges that are not components of the largest clique(s) the node belongs to.  To achieve this goal we propose a spectral technique that exploits the important observation made in the previous section: the magnitude of each entry of the dominant eigenvectors increases with the size of the clique that the corresponding node belongs to.

We will call a node that is not the original egocentric node, a sub-egocentric node, and define a sub-egonet to be the egonet for a sub-egocentric node restricted to the set of nodes and edges represented in the egonet under consideration; see Figure \ref{fig:SparsificationAlgorithmExplained_SubEgo} for an example.  To determine the largest clique(s) each sub-egocentric node belongs to, we extract its sub-egonet and approximate the dominant eigenvector of the corresponding adjacency matrix.  As larger values in this eigenvector correspond to nodes that are members of larger cliques, we remove all edges connecting the sub-egocentric node to a node whose entries in the dominant eigenvector is below half of the maximum for the eigenvector.  Figure \ref{alg:SparsifyEgo}, describes the egonet sparsification algorithm.

\begin{figure}[h]
\makebox{\SF Egonet Sparsification Algorithm}  \\
\rule[0mm]{\linewidth}{0.5pt} 
  \begin{itemize}
  \item [] {\SF Input}: 
    \begin{itemize}
    	\item node $v$
	\item  EGO($v$), the egonet of $v$
    \end{itemize}
  \item []{\SF Algorithm:}
	\begin{itemize}
	\item  Repeat steps 1 to 4 until the egonet's adjacency matrix no longer changes or a maximum number of iterations has been reached.
	\item  Repeat steps 1 to 3 for each node $u$ in EGO($v$).
	\end{itemize}
    \begin{enumerate}
    	\item  For each node, $u$, in EGO($v$) such that $u \neq v$, extract its sub-egonet, SEGO($u$). 
	\item  Scale the entries of the adjacency matrix for SEGO($u$) by $\delta = 1/N_{sego}$, where $N_{sego}$ is the number of nodes in SEGO($u$).  Call this matrix $A(u)$.
	\item  Approximate the dominant eigenvector of $A(u)$ using a small number of iterations, $O(10)$, of the power method.  If a node's index in the eigenvector has a magnitude below $1/2$ the maximum, remove the edge from EGO($v$).
	\item  Symmeterize EGO($v$) by removing any edges that became directed through the sparsification process.  
    \end{enumerate}
  \item  [] {\SF Output:}  Sparsified EGO($v$).
  \end{itemize}
  \rule[0mm]{\linewidth}{0.5pt} 
  \caption{Sparsification of egonet links
    \label{alg:SparsifyEgo}}
\end{figure}

Once the matrix has been sparsified, it can be analyzed using the spectral algorithm described in the previous section to generate the edge descriptor sets for the egocentric node.  The computational cost of this process is dominated by the cost of implementing the spectral algorithm, which is driven by the cost of calculating the eigenvectors of the sparsified ICM-matrix with positive eigenvalues.  Determining these eigenvectors costs $O(k^2)$ for a node of degree $k$ and thus $O(N~\overline{k^2})$ to generate them for an entire network involving $N$ nodes.

\subsubsection{Experimental Validation of the Sparsification Algorithm}

\begin{figure}[h!] 
 \begin{center}
	\includegraphics [width=0.45\textwidth] {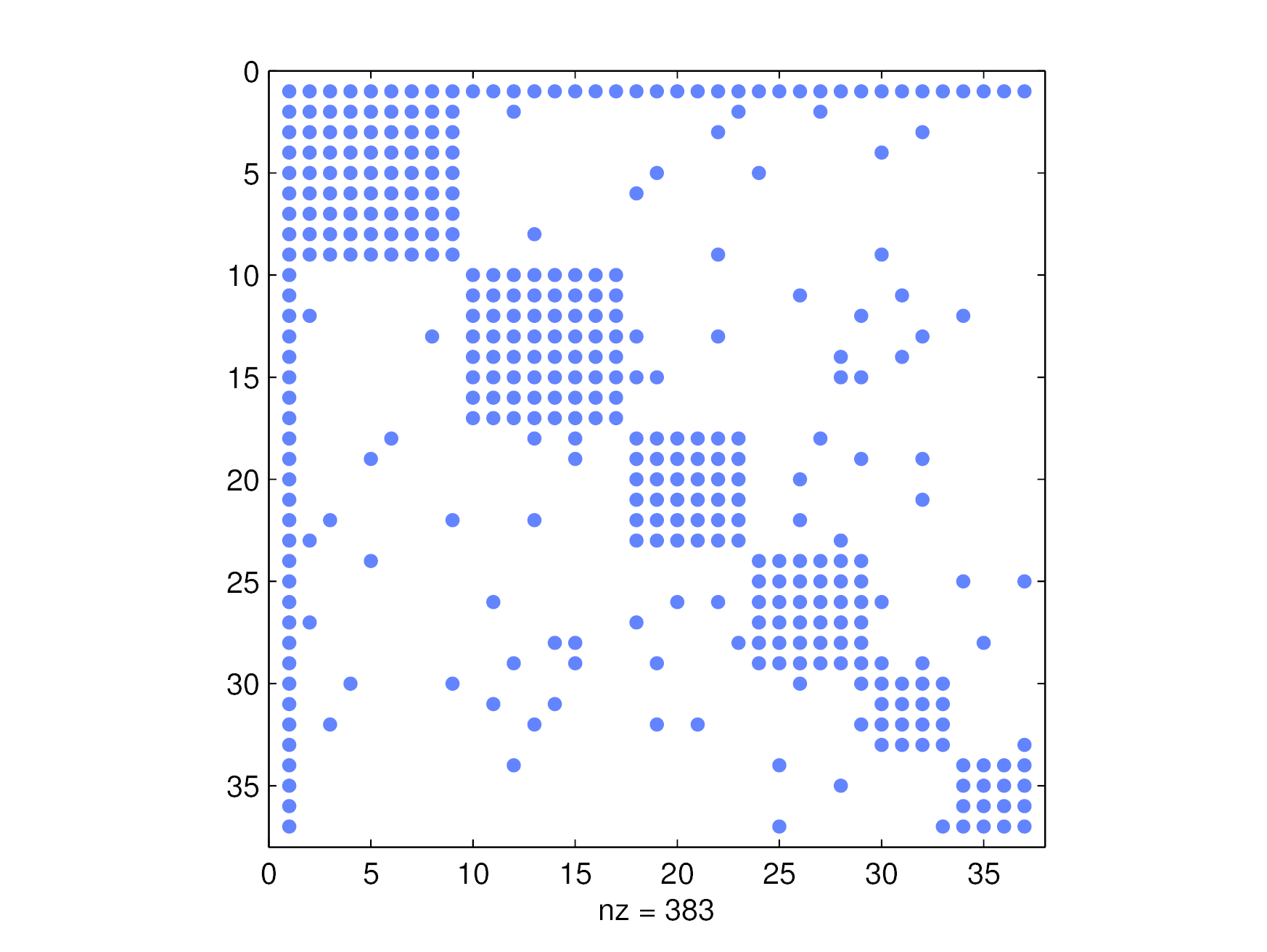} 
  \end{center}
   \caption{Example of an ICM-matrix with varying clique sizes and random connections added.  Each dot represents non-zero components of the matrix.}
\label{fig:CMatrixPreSparse}
\end{figure}

Our initial set of tests involve ICM-matrices that are perturbed by adding in random connections between nodes with probability $p$.  We use cliques composed of 4, 6, and 8 members, and plant two cliques of each size in the ICM matrix.  The value of $p$ is set according to Equation \eqref{eq:ChoiceOfP}, so that members of the smallest cliques have as many expected outlinks as they have inlinks.  

\begin{equation}
 	p = \frac{\text{Number~of~Nodes~in~Smallest~Clique}}{\text{Total~Number~of~Nodes~Outside~Smallest~Clique}}
\label{eq:ChoiceOfP}
\end{equation}•

 \noindent  An example of such matrices is shown in Figure \ref{fig:CMatrixPreSparse}.

Our findings show that the sparsification algorithm yields either an ICM-matrix or a small perturbation thereof when the random connections do not to alter the planted disjoint clique structure.  Even when the planted clique structure is altered through relatively large cliques developing via random connections, our findings are that the algorithm will still reliably recover the largest planted cliques.  

\begin{figure}[h!]
%  \begin{center}
    \subfigure [Pre-sparsification] { \label{fig:ICMatrixWin_PreSparse} \includegraphics [width=0.25\textwidth] {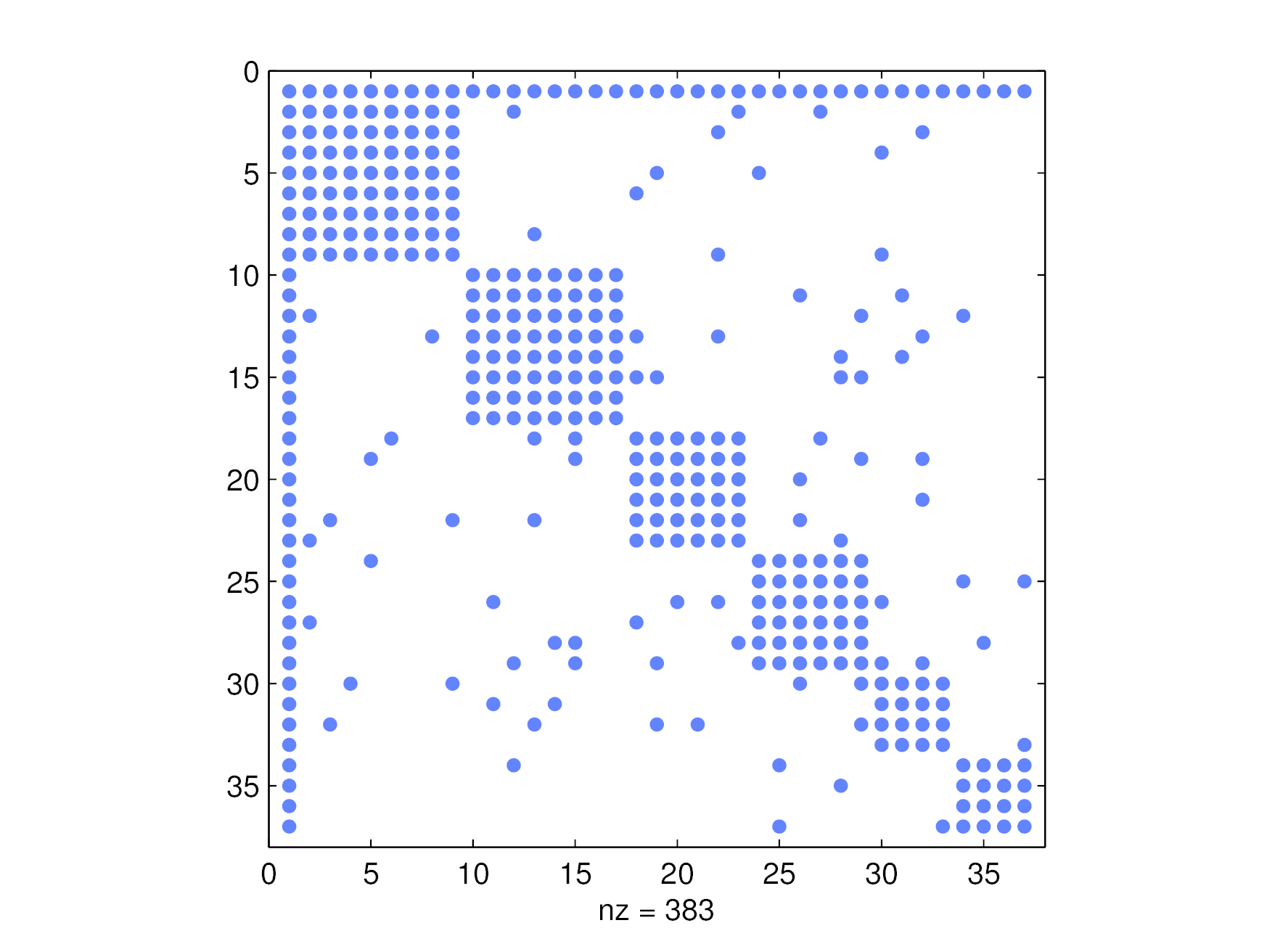} }%
    \subfigure [Post-sparsification] { \label{fig:ICMatrixWin_PostSparse} \includegraphics [width=0.25\textwidth] {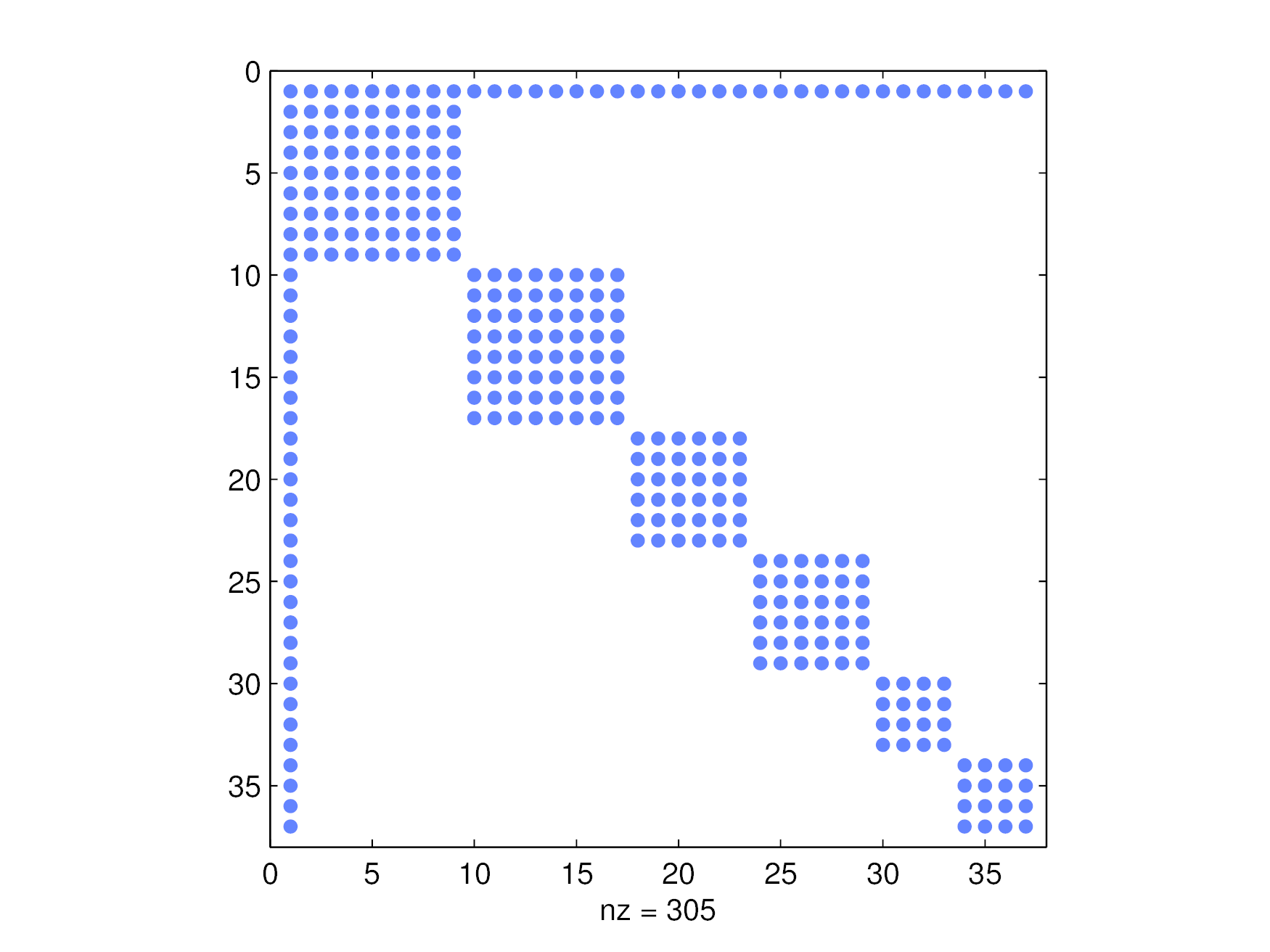} }%
%  \end{center}
   \caption{An example of the resulting matrix after a successful sparsification.  a) Matrix before a successful sparsification.    b) Matrix after a successful sparsification.  See text for discussion.}
  \label{fig:ICMatrixWin}
\end{figure}

\begin{figure}[h!]
%  \begin{center}
    \subfigure [ Pre-sparsification] { \label{fig:ICMatrixFail_PreSparse} \includegraphics [width=0.25\textwidth] {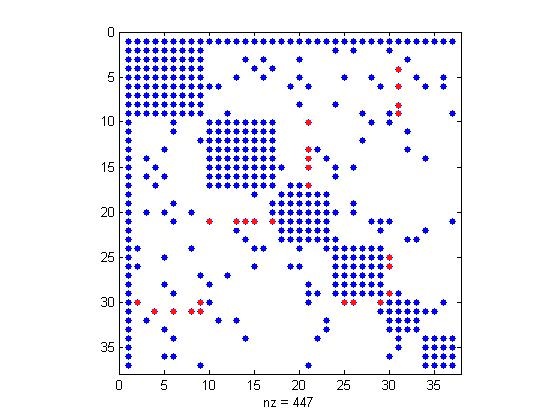} }%
    \subfigure [Post-sparsification] { \label{fig:ICMatrixFail_PostSparse} \includegraphics [width=0.25\textwidth] {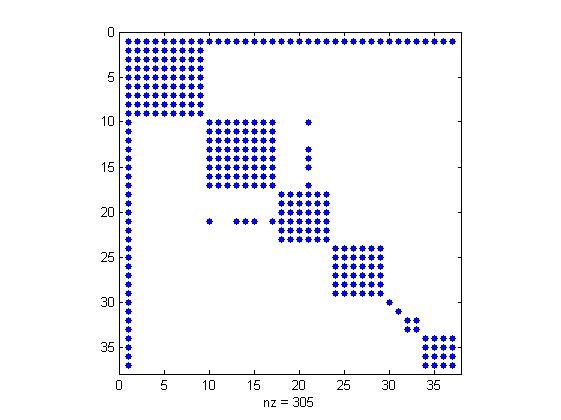} }%
%  \end{center}
   \caption{An example of the resulting matrix after a failed sparsification.  a) Matrix before an "unsuccessful" sparsification.    b) Matrix after an "unsuccessful" sparsification.  See text for discussion.}
  \label{fig:ICMatrixFail}
\end{figure}

Figure \ref{fig:ICMatrixWin} shows a typical example of a successful result: the algorithm correctly removes all of the random connections.  Figure \ref{fig:ICMatrixFail} exemplifies the cases where the algorithm fails to return the ideal planted partition.  Such cases correspond to a scenario where the random connections (colored in red in Figure \ref{fig:ICMatrixFail_PreSparse}) are generating cliques comparable in size to the original cliques that the nodes belong to, thus altering the clique structure that was originally planted.  When a node belongs to multiple cliques of approximately the same size, the connections will all remain so long as none overlap with a clique that is substantially (50\% or greater) larger in size, and the connections will all be pruned otherwise.  Both of these cases can be observed in Figure \ref{fig:ICMatrixFail_PostSparse}.  As the intention of the algorithm is to recover the largest disjoint cliques an egocentric node belongs to, neither of these outcomes are actually undesirable but they make determining the accuracy of the algorithm problematic.

%\pagebreak

%\clearpage

\begin{figure}[h]
%  \begin{center}
    \subfigure [ Five Groups of Ten] { 
	\begin{minipage}[c][0.75\width]{
	0.5\textwidth}
	\centering%
	\label{fig:FiveGroupsTenMembers} 
	\includegraphics [width=1\textwidth] {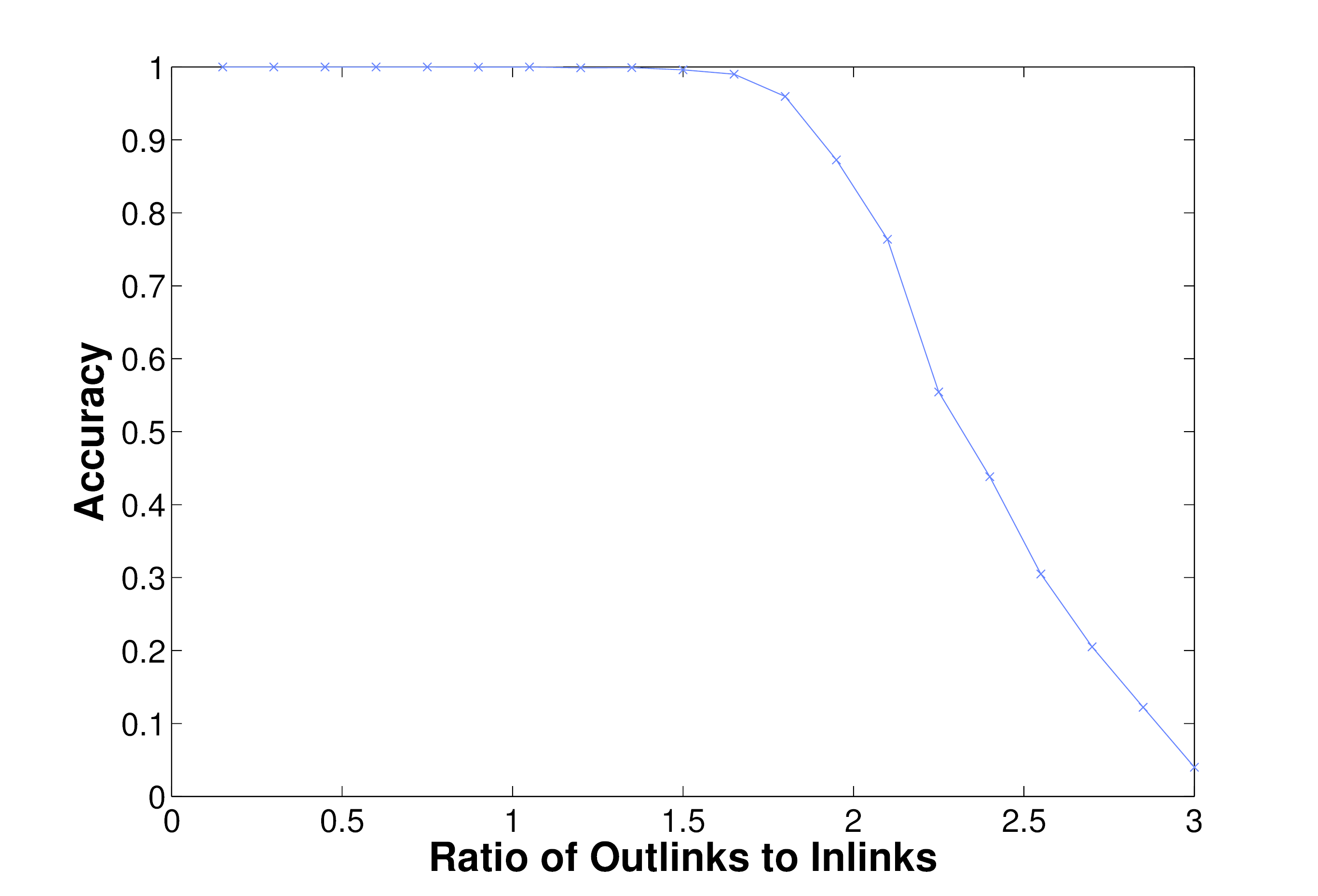} 
	\end{minipage}}
    \subfigure [ Ten Groups of Ten] {
	\begin{minipage}[c][0.75\width]{
	0.5\textwidth}
	\centering%
	 \label{fig:TenGroupsTenMembers}
	 \includegraphics [width=1\textwidth] {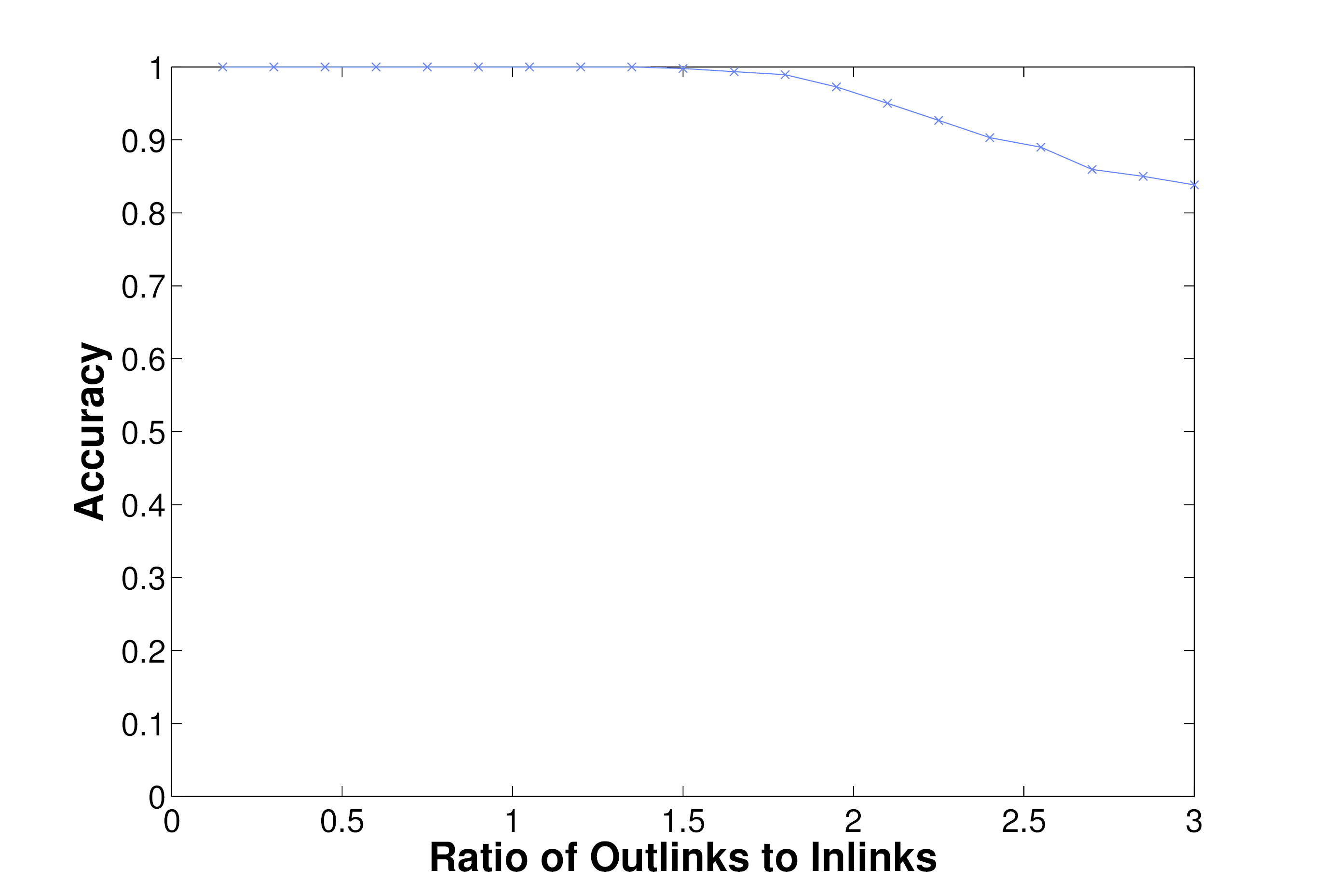}
	\end{minipage}}
%  \end{center}
   \caption{The connection removal accuracy of the algorithm, as a function of the expected ratio of the number of inlinks between a given node and the members of its planted clique and the number of outlinks between a given node and members of other planted cliques.  a)  This test is carried out on an ICM-matrix composed from five disjoint cliques with ten members each.  b)  The same test carried out in (a), but with the number of cliques doubled.}
  \label{fig:CMatrixPostSparse}
\end{figure}

To sidestep these difficulties, we tested the performance of the algorithm using cliques of fixed size (10 members) and increased the expected number of outlinks to other cliques from 0 to 30 (three times the number of inlinks).  We measured the accuracy of the sparsification algorithm by counting the percentage of properly removed random connections that were created during the perturbation process.  As can be seen in Figure \ref{fig:CMatrixPostSparse}, the algorithm performs remarkably well with 100\% accuracy, well past the point where there are more outlinks than inlinks in either case.

\subsection{Community Scale Features of Community Structure}
\label{sec:CommunityLevelPerspective}

\begin{figure}[h!] 
 \begin{center}
	\includegraphics [width=0.45\textwidth] {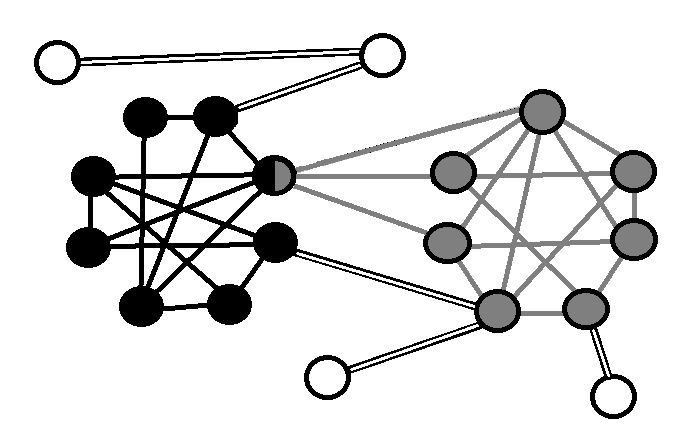} 
  \end{center}
   \caption{At the community level, the goal is to find collections of edge descriptor sets that form a community.  For the black and gray communities, we need to differentiate the sets of gray and black edges from the set of white edges. }
\label{fig:AlgorithmExplainedCommunityLevel}
\end{figure}

Once the edge descriptor sets have all been extracted, we need a way to stitch these localized features together to form communities.  To do this, we use quantitative features relevant to the scale of communities to agglomerate these edge sets.  

As is the case for the network depicted in Figure \ref{fig:AlgorithmExplainedCommunityLevel}, a reasonably general qualitative feature to expect communities to possess is that they have an edge density substantially higher than that of the network as a whole, and we note that this is not a property which is accurately represented at the level of individual nodes' edge descriptor sets.  To account for this property, we use link density in order to agglomerate edge descriptor sets to form communities.  

Let $C$ denote a set of nodes, and $E(C)$ denote the set of edges between all members of $C$. Given a set of $n$ vertices, the maximum number of edges is $n(n-1)/2$ (for a clique). We define the edge density as,

\begin{equation}
	\rho (C) = \frac{|E(C)|}{\frac{1}{2} |C| (|C| - 1)}.
\label{eq:LinkDensity}
\end{equation}

Our method for community formation is based on satisfying $\rho (C) = D$ where $0\leq D \leq 1$ is a user supplied inlink density and the number of nodes involved, $|C|$, is as large as possible.  The reason $D$ is taken as an input is because the appropriate value is dependent on the scale of the structures one is trying to extract from the network; lower threshold densities correspond to looser notions of what it means to be a community and higher threshold densities correspond to tighter ones.  

%Note that if $C$ represents a community being formed and we start with a clique as a community base for $C$, then $\rho (C)$ starts at its maximum possible value.  For any set of nodes we add to $C$, $\rho (C)$ can only decrease or stay the same, meaning it can be made to be a monotonically decreasing function as we successively add descriptor sets to the set defining the community.  

We take a greedy algorithmic approach to expanding communities, where the nodes involved in a new edge descriptor set are added to the community if they decrease $\rho$ by the minimal amount while still staying above the user supplied threshold.  Although this is not the most principled optimization approach from a mathematical standpoint, it preserves the intuitive notion of communities forming as a diffusive process of individual perspectives on the community structure of the network, similar to percolation of edge descriptor sets \citep{Pal05,FriendshipGroup10} or variations of label passing \citep{Raghavan07, GraphSwarm12}.  An outline of the community expansion algorithm is given in Figure \ref{alg:CommunityFormation}.

\begin{figure}[h]
\makebox{\SF Community Formation Algorithm}  \\
\rule[0mm]{\linewidth}{0.5pt} 
  \begin{itemize}
  \item [] {\SF Input}: 
    \begin{itemize}
    	\item All edge descriptor sets detected on the network.
    \end{itemize}
  \item []{\SF Algorithm:}
    \begin{enumerate}
	\item  Start with largest edge descriptor set that remains unclustered as a community base.
    	\item  Find the edge descriptor set that would cause the minimum reduction in inlink density for the community being formed.
	\item  If the inlink density would remain above the user supplied density threshold, add the descriptor set to the community being formed.
	\item  Repeat Steps 2 and 3 until no edge descriptor set satisfies the density constraint.
	\item  Repeat from Step 1 until no edge descriptor sets remain unclustered.	
    \end{enumerate}
  \item  [] {\SF Output:}  Initial set of communities.
  \end{itemize}
  \rule[0mm]{\linewidth}{0.5pt} 
  \caption{Sparsification of egonet links
    \label{alg:CommunityFormation}}
\end{figure}

We conclude this section with a brief cost analysis of the community formation process.  Let $S$ be the set of nodes involved in a potential edge descriptor set to add to a forming community, $C$.  Let $N_C$ represent the total number of nodes that either belong to $C$ or are connected to one of the nodes in $C$ at any given stage of the community formation process.  Each update to $\rho(C)$ involves a sweep over all edge descriptor sets for each of the $N_C$ nodes.  Adding a potential edge descriptor set involves checking the density of the original adjacency matrix for the original network restricted to the indices representing nodes involved in the potential updated version of $C$.  Because this resulting density can be written in terms of a sum of the link density before the addition and the link density of the addition, the only substantial computational cost to check a potential update comes from calculating the density of the addition, a calculation that has a computational cost of $O(N_C)$ flops so long as $|S| << N_C$.  As this must be carried out for each of the $N_C$ nodes involved in $C$, the cost of each update is proportional to $N_C^2$.  This implies that the computational cost to extract the entire community is bounded above by a constant multiple of $N_{com}^2$, where $N_{com}$ is the total number of nodes involved in that community.  Since this must be carried out for each community in the network, the total computational cost for the community formation process is proportional to $O(\overline{N_{com}^2})$, the average of the square of community sizes.

\subsection{Network Scale Features of Community Structure}
\label{sec:NetworkLevelPerspective}

Lastly, we must address the question of what properties a collection of communities should have at the level of the entire network.  The answer to this question depends on the expectations of the researcher applying the algorithm, but as general heuristics we will require that there are no unclustered nodes in the network and that only the minimal required number of communities will be returned which provide a cover for the set of nodes in the network.  

% [[  NOTE:  CHECK TO SEE WHAT IS GOING ON WITH [b], IT JUST SENDS EVERYTHING TO THE END OF THE PAPER ]]

\begin{figure}[h] 
 \begin{center}
	\includegraphics [width=0.45\textwidth] {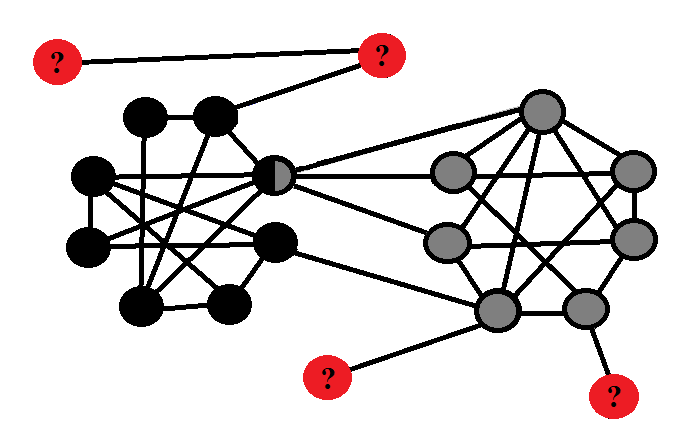} 
  \end{center}
   \caption{ At the network level, the goal is to make sure that the collection of communities returned by the algorithm satisfy criteria expected by the researchers using it.  For the network depicted above, a natural question is what to do with the nodes that do not decisively belong to any communities detected.}
\label{fig:AlgorithmExplainedNetworkLevel}
\end{figure}

Given the initial set of communities formed using the method described in the preceding section, we force all nodes to belong to at least one community by taking any nodes that remain unclustered after the community formation process and assign them to the community they share the most edges with.  This process is carried out iteratively if needed.  We then build a cover for the network out of the communities detected, starting with the largest as an initial element in the set.  Communities are then successively added to the set comprising the cover based on having the lowest percentage overlap with the current cover.  When multiple communities are tied for the lowest percentage, larger communities are given preference over smaller ones.  This process is carried out until every node in the network is represented in at least one of the communities in the set.

\section{Experiments}

We validate our approach on four different datasets. We first consider synthetically generated networks, coming from the planted $l$-partition and LFR benchmark tests.  The other two datasets are based on graphs from real world social networks, one the famous Zachary Karate Club network \citep{Zachary77} and the other a high school friendship network \citep{Xi13}.  

We view the task of community detection as querying a given network for the community structures present, therefore we use F-score to assess the quality of our community detection algorithm.   The F-score of a result is a common measure to use in gauging the quality of information retrieval applications, and is defined by Equations \eqref{eq:Precision} through \eqref{eq:Fscore}.  

\begin{equation}
\resizebox{1.0 \hsize}{!} 	{$Precision =  \frac{| \{\text{Gold~Standard~Community} \} \cap \{\text{Detected~Community} \} | }{ | \{\text{Detected~Community} \}  | }  .$}
\label{eq:Precision}
\end{equation}•

\begin{equation}
 \resizebox{1.0 \hsize}{!} 	{$	Recall =  \frac{| \{\text{Gold~Standard~Community} \} \cap \{\text{Detected~Community} \} | }{ | \{\text{Gold~Standard~Community} \}  | }   . $}
\label{eq:Recall}
\end{equation}•

\begin{equation}
 	F =  \frac{2 \times \text{Precision} \times \text{Recall}}{\text{Precision} + \text{Recall}}  .
\label{eq:Fscore}
\end{equation}•

The elements of each set are taken as the nodes involved in the community, and the precision and recall values we report are taken as the average precision and recall values when each planted community is paired with the detected community with the highest F-score.  The benefit of using this metric to assess our algorithm is that the precision scores reflect the quality of our choice for edge descriptor sets, and the recall scores reflect the quality of our choices for community formation and desired network level properties.

For the high school and LFR tets, we additionally calculate the extended notion of normalized mutual information (NMI) \citep{La09,La09a} between the two sets of community labels in order to measure the algorithm's performance.  The motivation for including this measure for these tests is that this was the performance metric used in \citep{Xi13}, and allows one to roughly compare our algorithm to a host of others.  Although the extended notion of normalized mutual information slightly differs from what is standardly called NMI, we will refer to this extended version as just NMI in this paper for the sake of simplicity.  An NMI score of 1 implies that the labels correlate perfectly, and an NMI score of zero indicates that the labels have no correlation with each other.  We refer the reader to \citep{La09} for a detailed technical description of normalized mutual information.

\subsection{Planted $l$-Partition Benchmark Tests}

Our first set of synthetic network tests are planted $l$-partition tests.  These are standard benchmark tests \citep{GN2001} that create randomly generated networks with planted communities, where nodes in the same community have a higher probability to be connected than nodes in differing communities.   The test involves fixing the expected degree of a node, and increasing the expected proportion of those links which are outlinks to other communities.  Not only does this make the boundary between communities less well defined because there are more links between communities, but it also makes communities less well defined by decreasing their inlink density.  Because our algorithm is built to detect communities that can overlap, we again use F-scores to measure the quality of the results instead of the more standard metric of recovering the planted partitioning of nodes.  The motivation for using F-scores is that there is not a one to one correspondence between sets of correctly partitioned nodes and correctly identified communities when it is assumed that communities can overlap with one another.

\begin{figure}[h!]
  \begin{center}
     \includegraphics [width=0.5\textwidth] {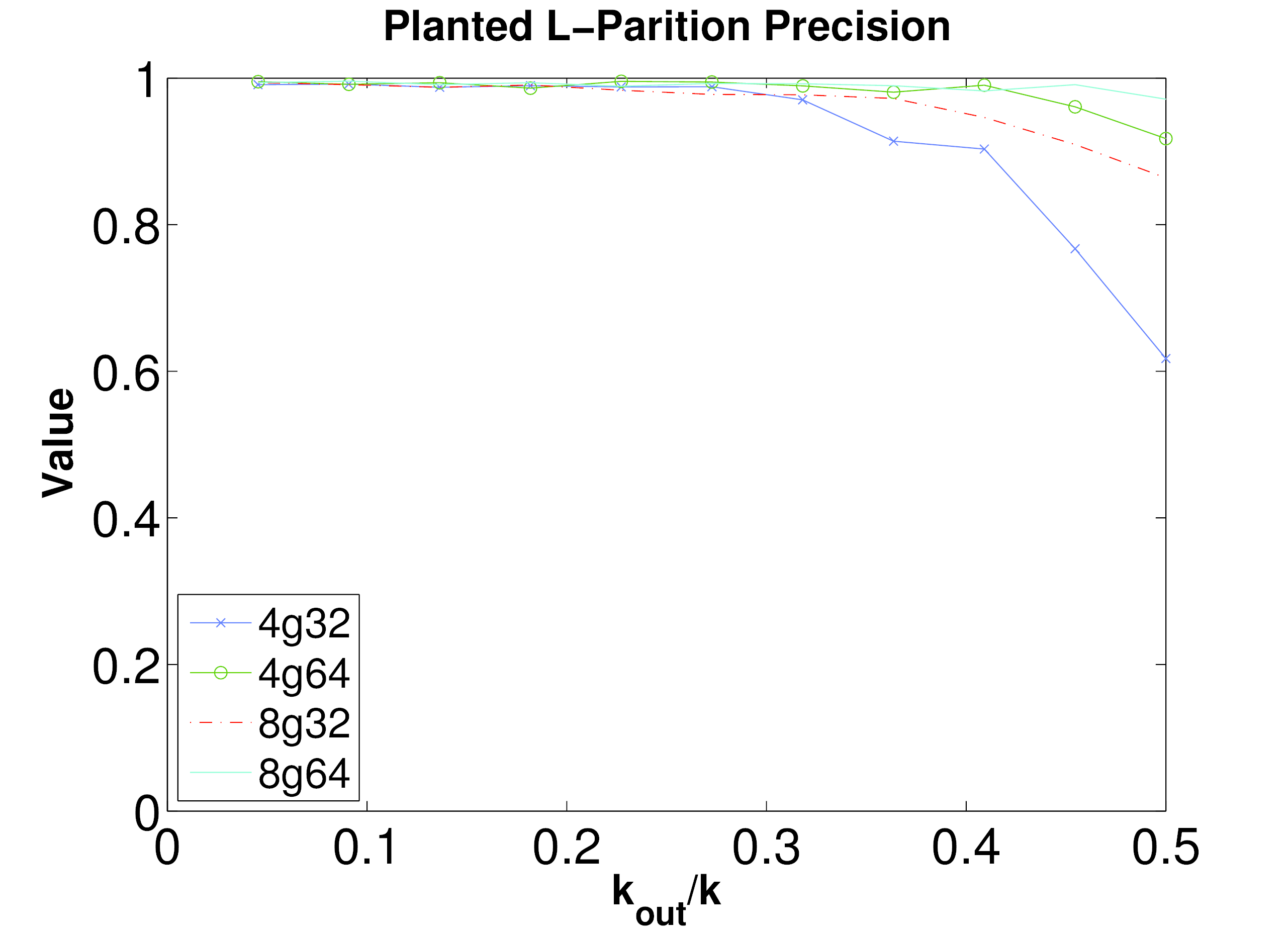} %
  \end{center}
   \caption{  Precision scores for the planted $l$-partition tests.}
  \label{fig:PL_Precision}
\end{figure}

\begin{figure}[h!]
  \begin{center}
    \includegraphics [width=0.50\textwidth] {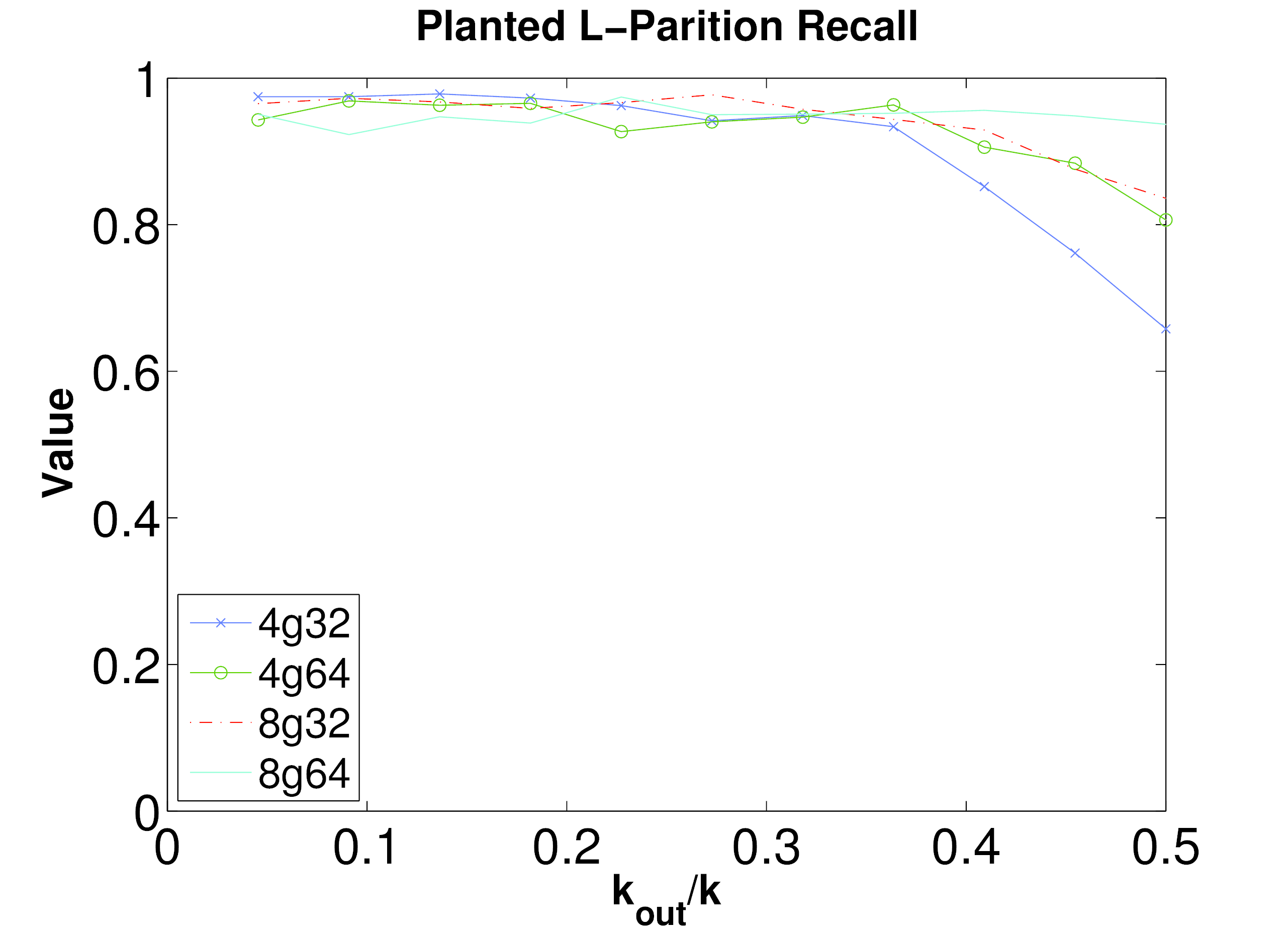} %
  \end{center}
   \caption{  Recall scores for the planted $l$-partition tests.}
  \label{fig:PL_Recall}
\end{figure}

\begin{figure}[h!]
  \begin{center}
     \includegraphics [width=0.50\textwidth] {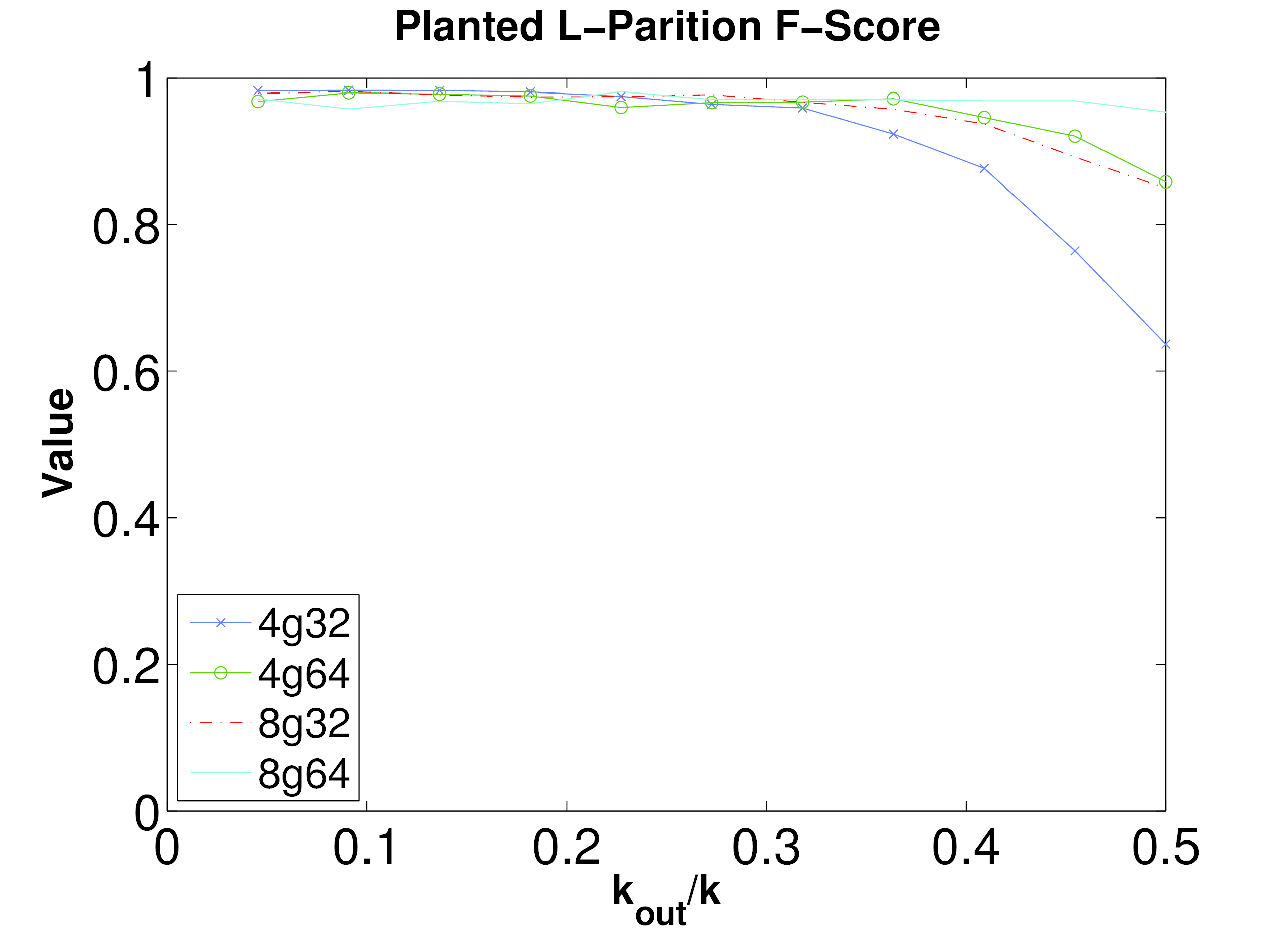} %
  \end{center}
   \caption{   F-scores for the planted $l$-partition tests.  }
  \label{fig:PL_F}
\end{figure}

We conduct several versions of the planted $l$-partition test:  four groups of 32 members, eight groups of 32 members, four groups of 64 members, and eight groups of 64 members.  For all of these tests, the expected degree per node is set to be equal to half of the total number of members in each planted community.  The first test with four groups of 32 is the most standard, and allows one to roughly compare our algorithm against a host of others \citep{DaEtAl05}.  The remaining tests provide a controlled setting to demonstrate how the algorithm's performance substantially improves when there are more communities and/or larger communities, which more accurately reflects the types of networks the algorithm was intended for.  The precision, recall, and F-scores for this series of tests are presented in Figures \ref{fig:PL_Precision}-\ref{fig:PL_F}, where each data point represents the averaged result computed over twenty independent realizations of the random network.  Each test is denoted by [Number of Groups]g[Number of Members per Group].  Note that the algorithm's performance increases in all respects with either increased community sizes or number of communities, and the F-scores for the test involving eight groups of 64 still remains above 0.95 even in the case where half of the links for any node are expected to be outlinks.

%\pagebreak

\subsection{LFR Benchmark Tests}

The second set of synthetic network tests consists of the  LFR benchmark tests  \citep{La09b} which are designed to construct synthetic networks with built-in community structures.  LFR networks are a variation of the planted $l$-partition model, where nodes are no longer required to all have the same expected degree and communities can come in varying sizes.  Although more general versions of the test exist in which edge weights and directions are also considered, we only focused on the extended version of the test that allows for nodes to be members of multiple communities. Nodes belonging to multiple communities will be referred to as \textit{overlapping nodes}, where the total number of overlapping nodes in a given network is referred to as $O_n$ and the number of communities an overlapping node belongs to as $O_m$.

In order to be able to compare the results of our algorithm with the experiments conducted in the recent survey \citep{Xi13}, we use the same parameters as \citep{Xi13} for many of our experiments. Node degree and community sizes are respectively drawn from power law distributions with $\tau_1 = 2$ and $\tau_2 = 1$, the average degree per node is set to $k_{ave} = 10$, and the maximum degree a node can have is set to $k_{max} = 50$.  The remaining parameters were varied throughout the tests.  We use networks with sizes $N \in \{1000,5000\}$, and community sizes in both a small range $s=(10,50)$ and a large range $b=(20,100)$.  The fraction of links through which a node connects to members of other communities is denoted by the mixing parameter $\mu$.  As with the planted $l$-partition tests, we randomly generate 20 networks for each set of parameter values and report the average performance.

In the first set of experiments, we keep the maximum number of overlapping communities constant, $O_m =2$. We increase the density of edges between communities: $\mu$ is increased from 0.1 to 0.3 by increments of 0.05.  All of the combinations of parameter values for $N$ and community sizes are examined, along with setting $O_n$ to either 10\% or 50\% of the nodes in the network.  For this set of tests, we examine the average precision, recall, F-score, and NMI for each set of communities returned by the algorithm compared against those planted by the test.  The cut-off density for community expansion is set to $(1-\mu)$ multiplied by the average egonet density of the graph.

\begin{figure}[h!]
  \begin{center}
    \includegraphics [width=0.5\textwidth] {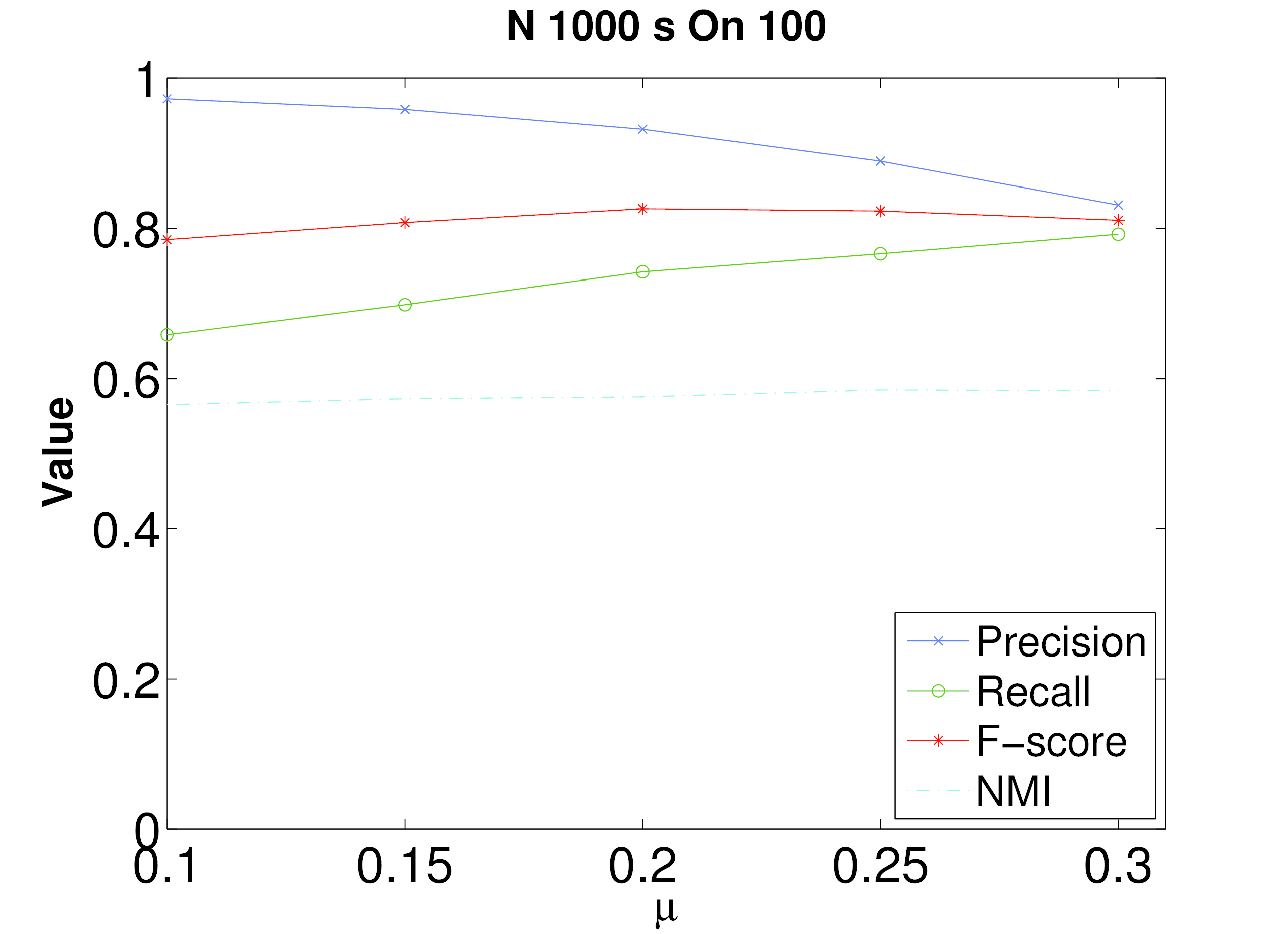} 

  \end{center}
   \caption{  The precision, recall, F-score, and NMI for the LFR tests varying $\mu$ for a network with 1000 nodes and using the smaller community size range.  $O_n = 10\%$ of the total nodes.  }
  \label{fig:VaryMu_N1000s}
\end{figure}

\begin{figure}[h!]
  \begin{center}
     \includegraphics [width=0.5\textwidth] {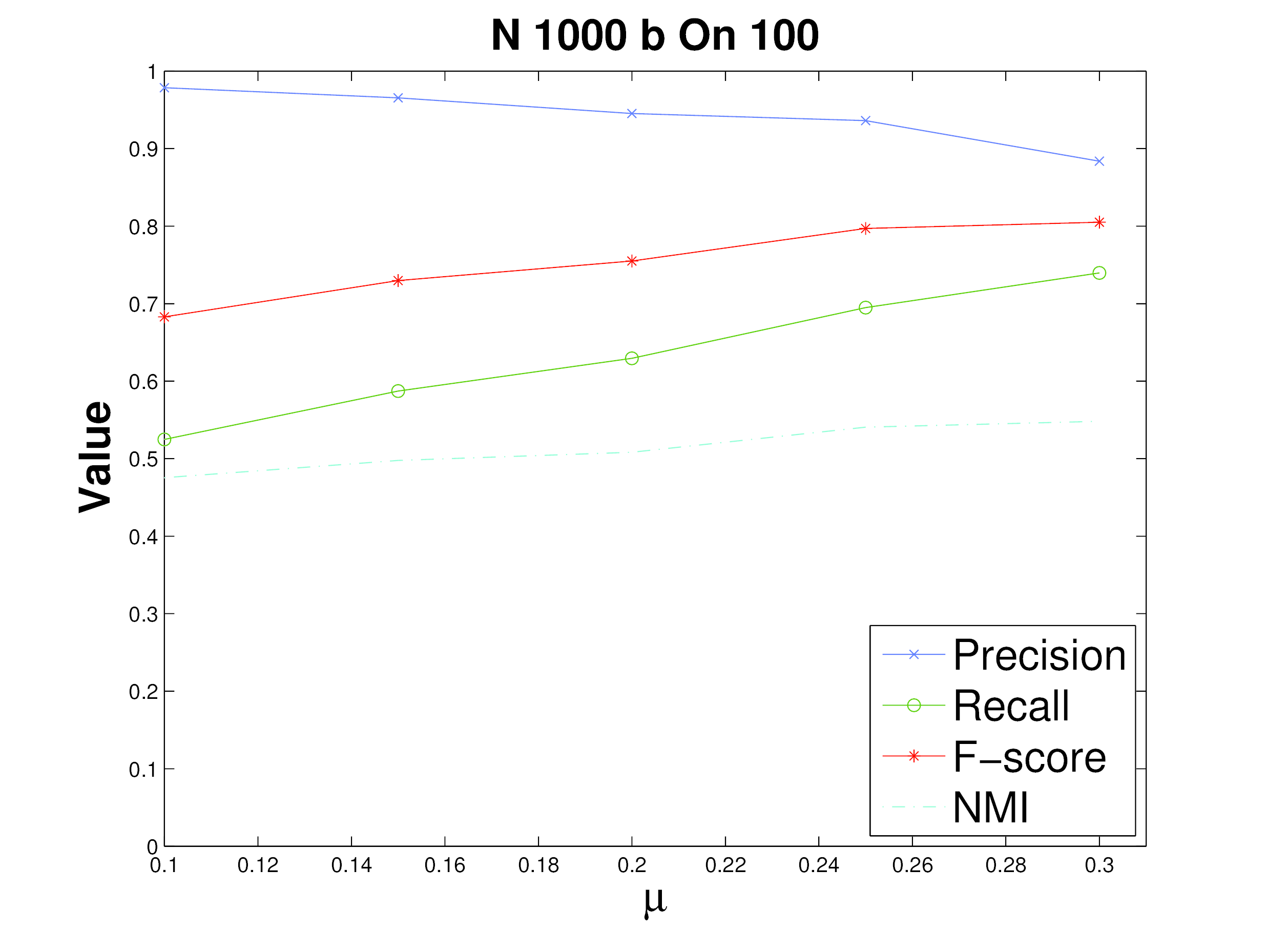} 
  \end{center}
   \caption{  The precision, recall, F-score, and NMI for the LFR tests varying $\mu$ for a network with 1000 nodes and using the larger community size range.  $O_n$ = 10\% of the total nodes.  }
  \label{fig:VaryMu_N1000b}
\end{figure}

\begin{figure}[h!]
  \begin{center}
    \includegraphics [width=0.5\textwidth] {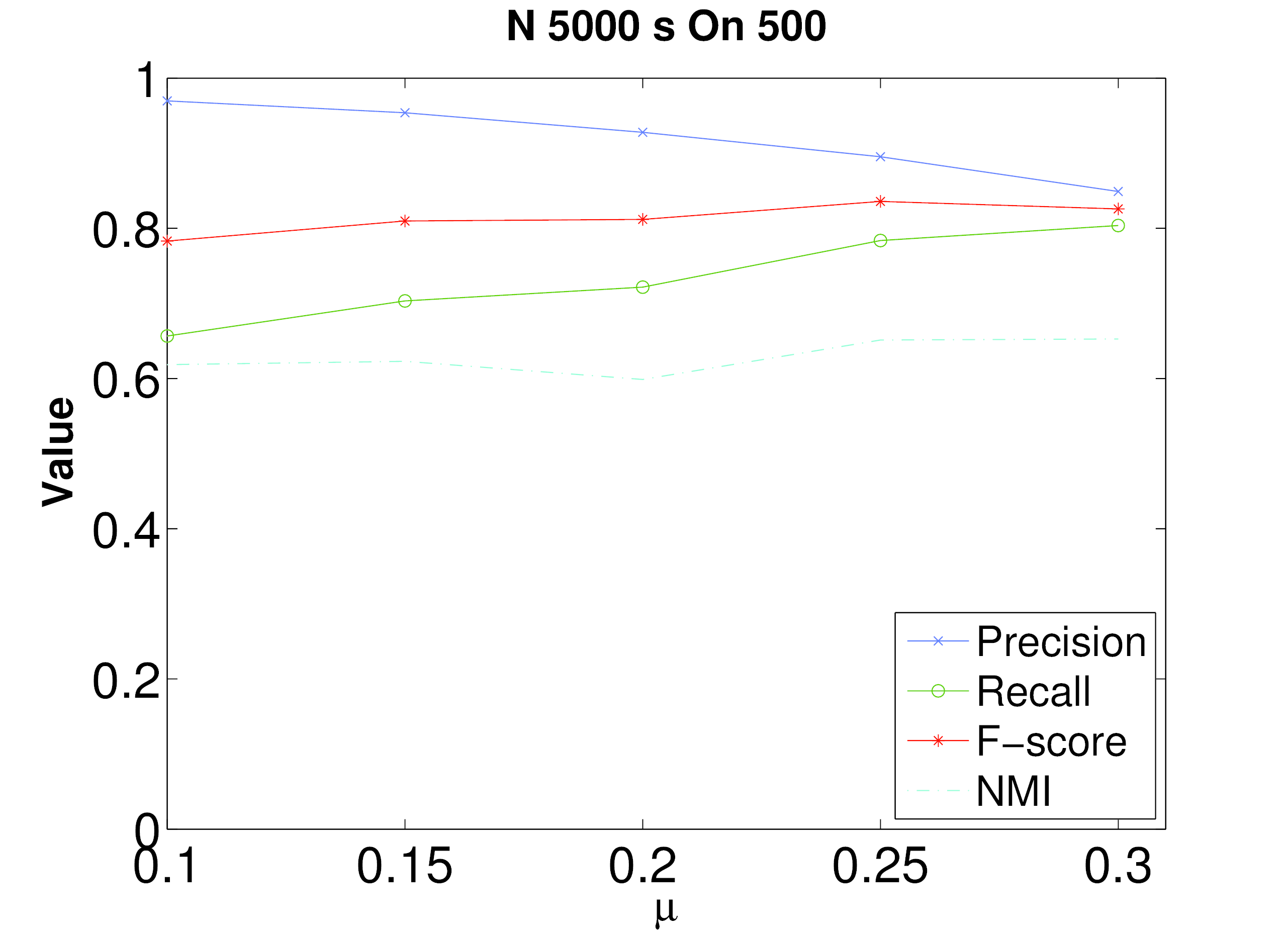} 
  \end{center}
   \caption{  The precision, recall, F-score, and NMI for the LFR tests varying $\mu$ for a network with 5000 nodes and using the smaller community size range.   $O_n$ = 10\% of the total nodes.   }
  \label{fig:VaryMu_N5000s}
\end{figure}

\begin{figure}[h!]
  \begin{center}
     \includegraphics [width=0.5\textwidth] {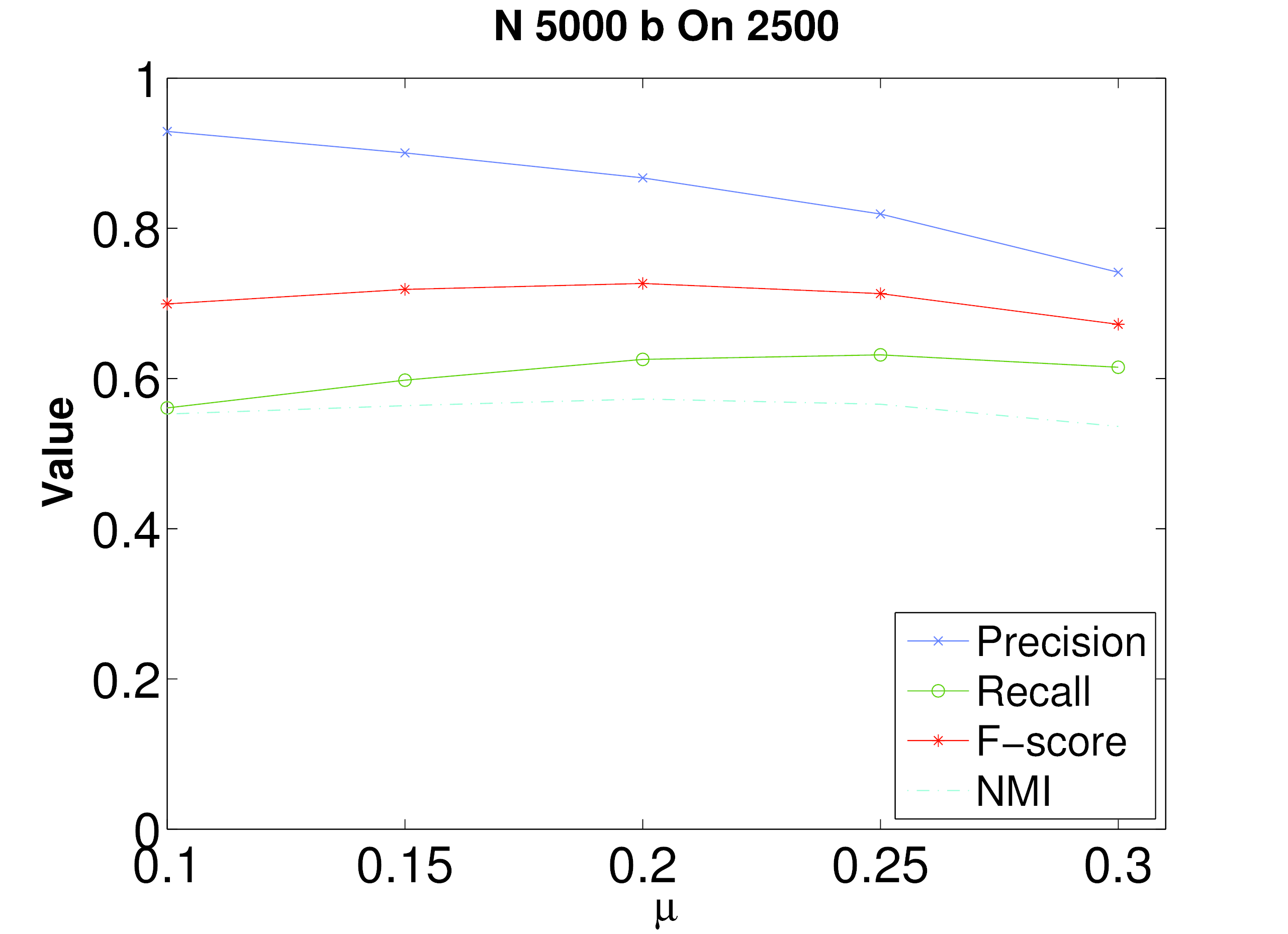} 
  \end{center}
   \caption{  The precision, recall, F-score, and NMI for the LFR tests varying $\mu$ for a network with 5000 nodes and using the larger community size range.  $O_n$ = 50\% of the total nodes.  }
  \label{fig:VaryMu_N5000b}
\end{figure}

As we can see from Figures \ref{fig:VaryMu_N1000s}-\ref{fig:VaryMu_N5000b}, the algorithm's performance again improves as the number and the sizes of the communities increase.  Somewhat surprisingly, the algorithm's F-scores also tends to increase with increasing values of the mixing parameter for the low overlap cases, where $O_n = 10\%$ of the total nodes.  This apparent paradox can be explained by observing that increasing $\mu$ lowers the edge density within communities. We can use a lower edge density threshold to merge the cliques, and recover the communities, and the recall score is improved.  Also, because $O_n$ is small, communities still remain well separated, and spurious cliques are not created by the increase in outlinks. However, for the high overlap case, we find that increasing the value of the mixing parameter tends to have only minor effects on the recall scores while significantly impairing the accuracy scores.

\begin{figure}[h!]
  \begin{center}
   \includegraphics [width=0.5\textwidth] {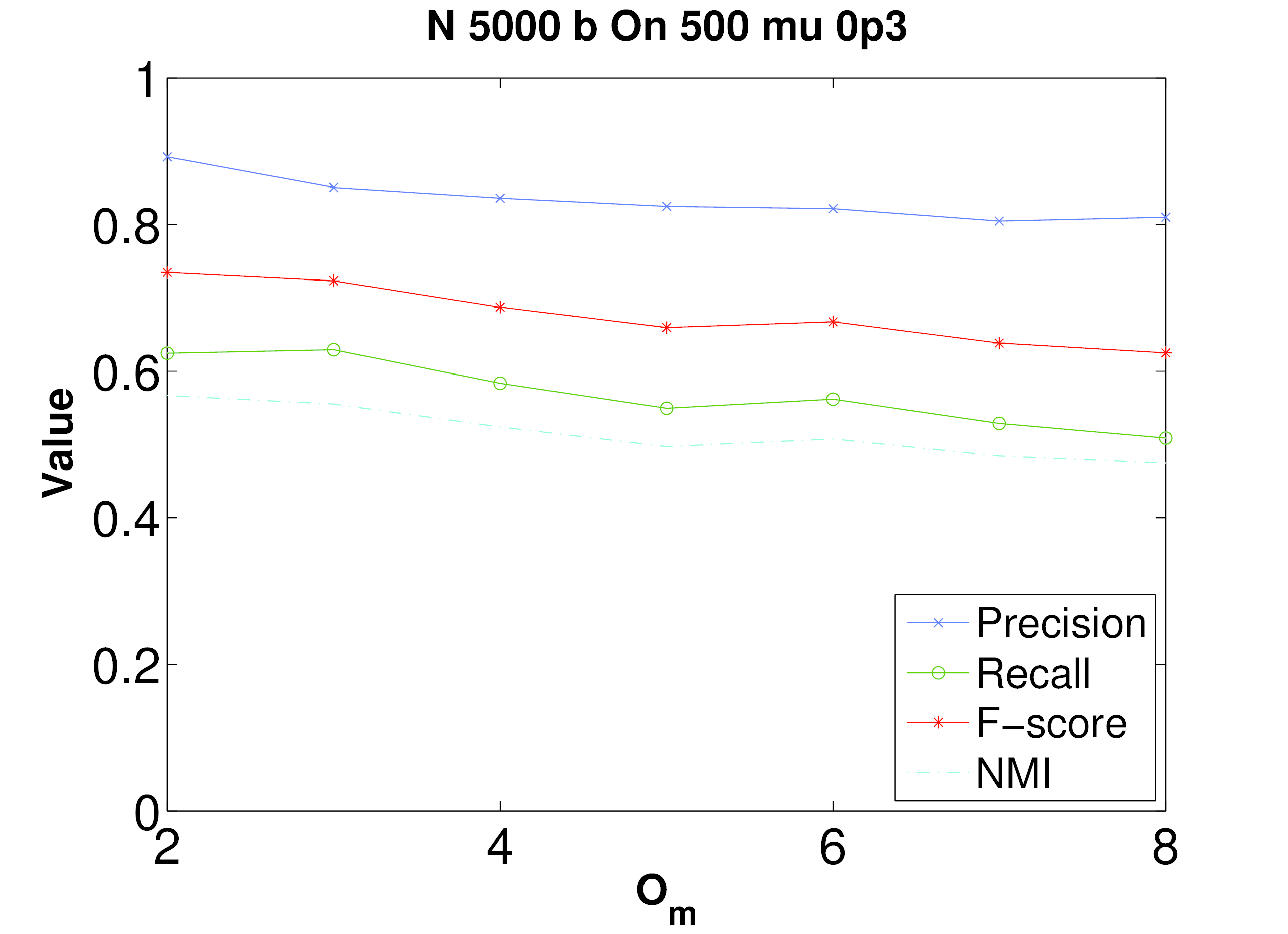} 
  \end{center}
   \caption{  The precision, recall, F-score, and NMI for the LFR tests varying $O_m$ for networks with 5000 nodes with 10\% of the total nodes belonging to two different communities, using large community size range distribution.  }
  \label{fig:VaryOm_N5000}
\end{figure}

The second subset of tests varies $O_m$ from 2 to 8, with $N=5000$, $\mu=0.3$, and $O_n = 10\%$.  This set of tests were also conducted on a variety of overlapping community detection algorithms in \citep{Xi13}, but only the NMI was examined in that work.  The precision, recall, F-scores, and NMI of our algorithm for this set of tests are presented in Figure \ref{fig:VaryOm_N5000}.  Although the NMI of our algorithm on this series of tests is about average with respect to all the algorithms analyzed in \citep{Xi13}, our precision scores are excellent.

%-----------------------------------------------------

\begin{figure}[h!]
  \begin{center}
     \includegraphics [width=0.5\textwidth] {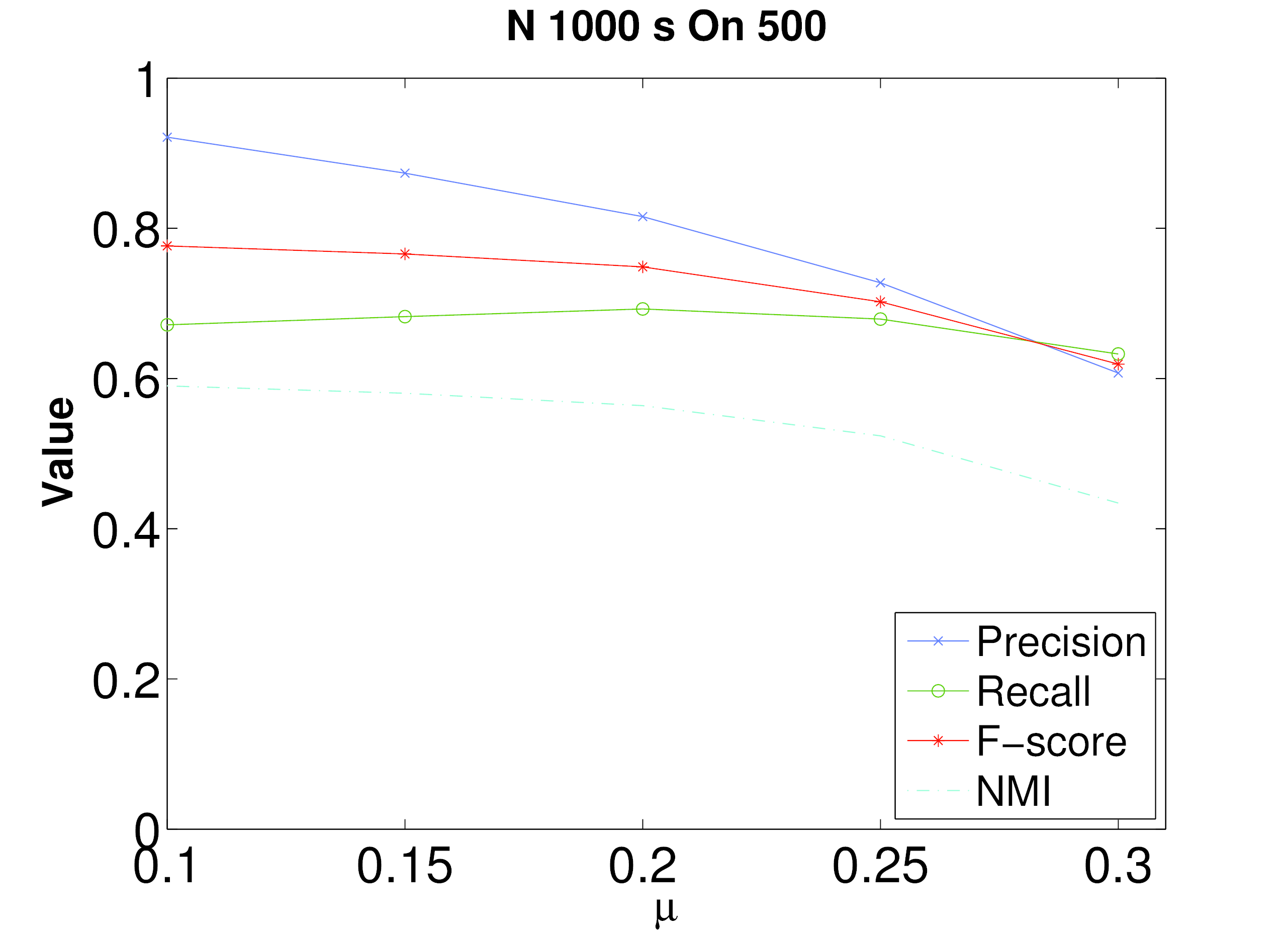} 
  \end{center}
   \caption{  The precision, recall, F-score, and NMI for the LFR tests varying $\mu$ for a network with 1000 nodes and using the smaller community size range.  $O_n = 50\%$ of the total nodes.  }
%  \label{fig:VaryMu_N1000s}
\end{figure}

%-----------------------------------------------------

\begin{figure}[h!]
  \begin{center}
    \includegraphics [width=0.5\textwidth] {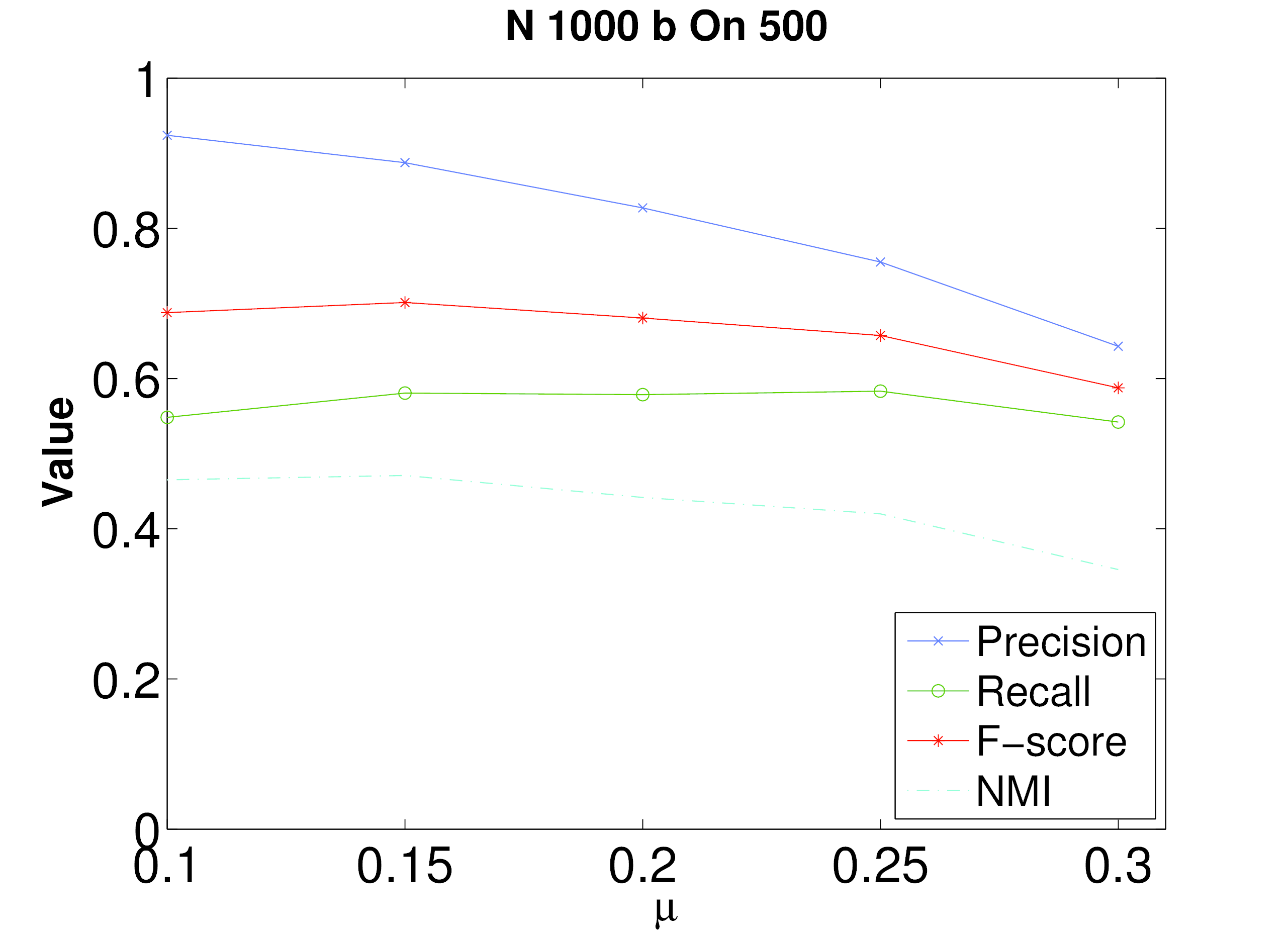} 
  \end{center}
   \caption{  The precision, recall, F-score, and NMI for the LFR tests varying $\mu$ for a network with 1000 nodes and using the larger community size range.  $O_n$ = 50\% of the total nodes.  }
%  \label{fig:VaryMu_N1000b}
\end{figure}

%-----------------------------------------------------

\begin{figure}[h!]
  \begin{center}
     \includegraphics [width=0.5\textwidth] {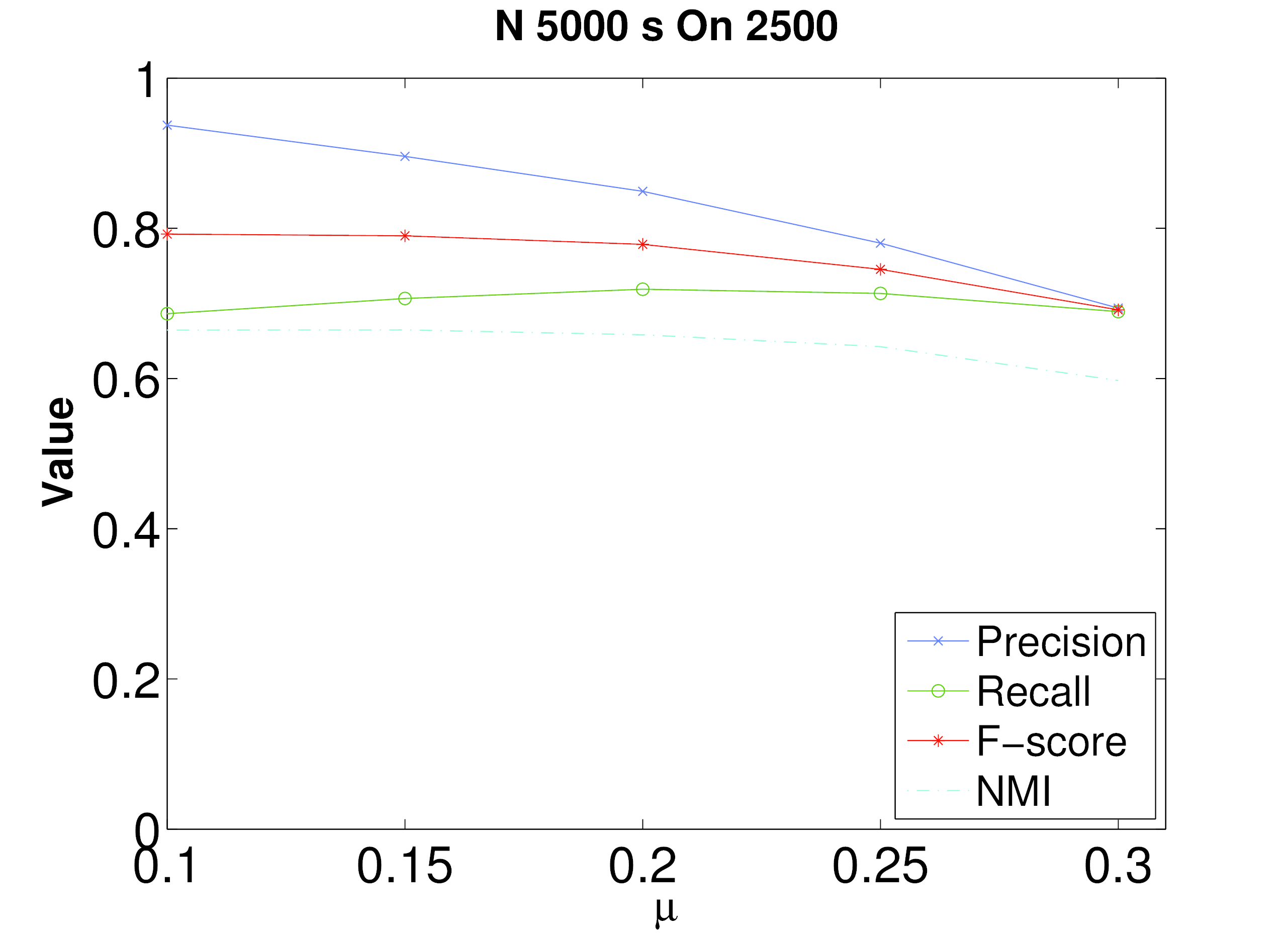} 
  \end{center}
   \caption{  The precision, recall, F-score, and NMI for the LFR tests varying $\mu$ for a network with 5000 nodes and using the smaller community size range.   $O_n$ = 50\% of the total nodes.  }
%  \label{fig:VaryMu_N5000s}
\end{figure}

%-----------------------------------------------------

\begin{figure}[h!]
  \begin{center}
     \includegraphics [width=0.5\textwidth] {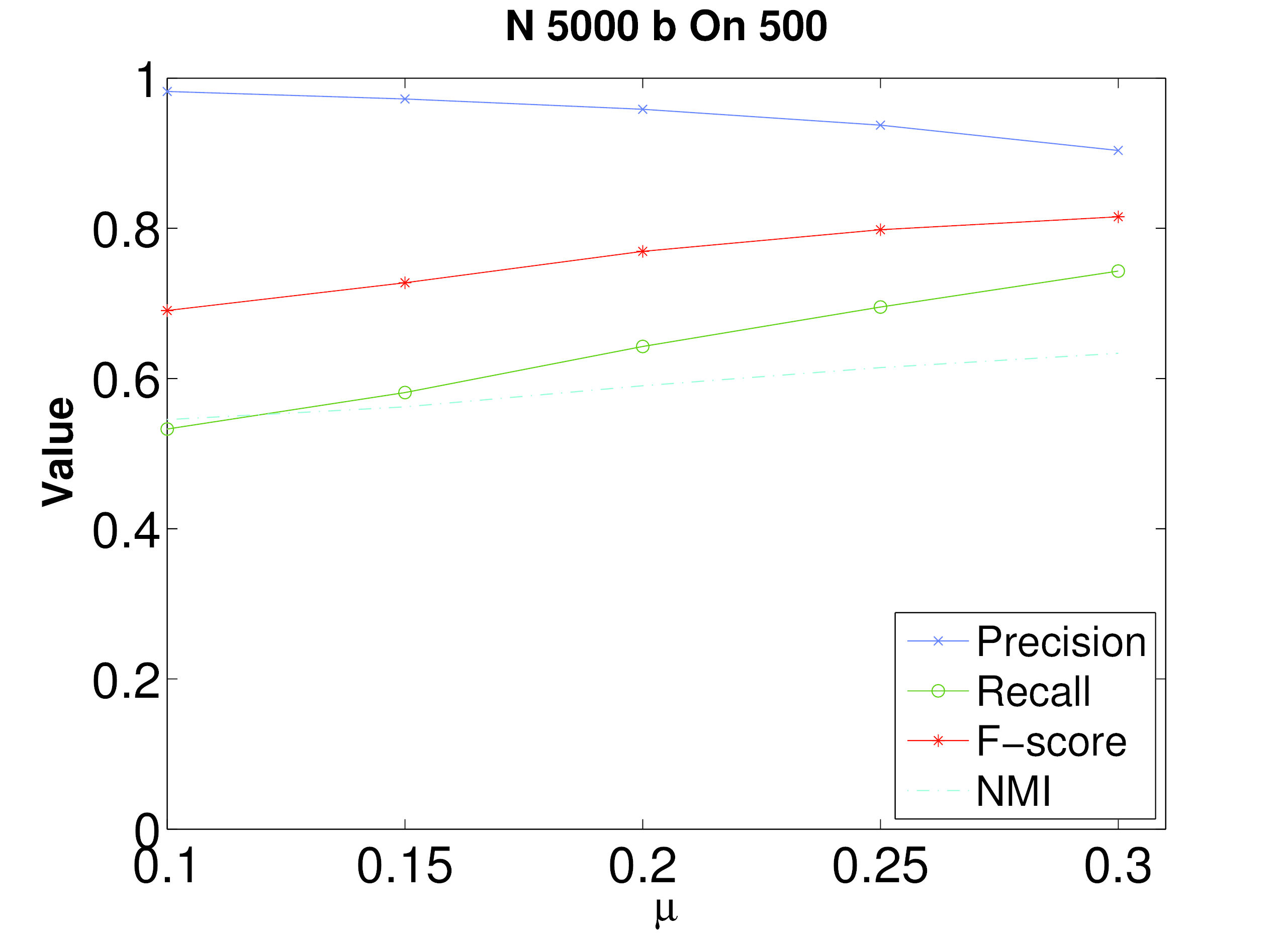} 
  \end{center}
   \caption{  The precision, recall, F-score, and NMI for the LFR tests varying $\mu$ for a network with 5000 nodes and using the larger community size range.  $O_n$ = 10\% of the total nodes.    }
%  \label{fig:VaryMu_N5000b}
\end{figure}

%-----------------------------------------------------

\begin{figure}[h!]
  \begin{center}
    \includegraphics [width=0.5\textwidth] {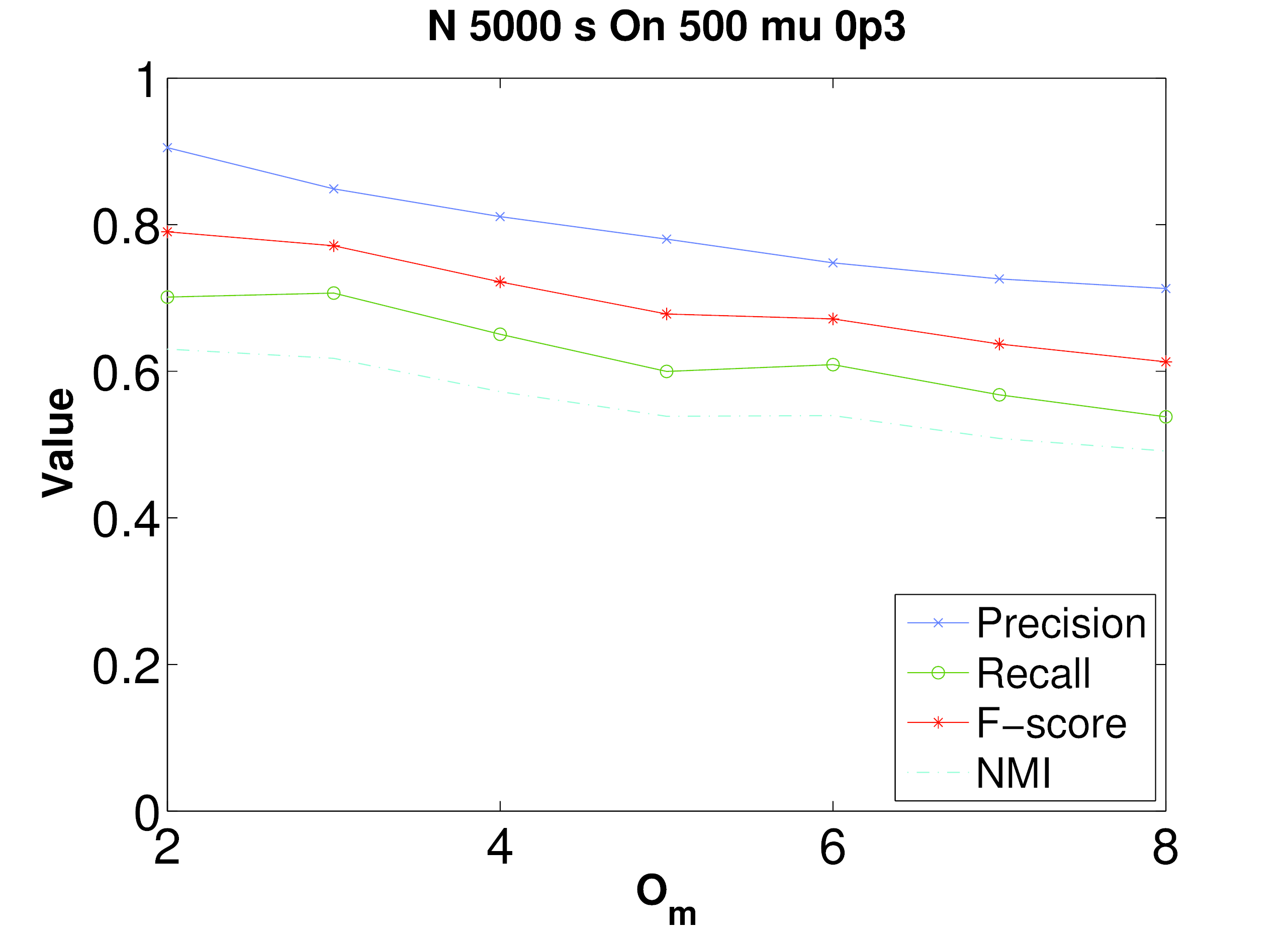} 
  \end{center}
   \caption{  The precision, recall, F-score, and NMI for the LFR tests varying $O_m$ for networks with 5000 nodes with 10\% of the total nodes belonging to two different communities, using small community size range distribution.  }
%  \label{fig:VaryOm_N5000}
\end{figure}

%\pagebreak

\subsection{Zachary Karate Club}
\label{sec:Karate}

Zachary's karate club network is a small social network comprised of the interactions amongst members of a university karate club studied by sociologist Wayne Zachary in the 1970's \citep{Zachary77}.  During the period of study, a political issue arose regarding the club's fees which eventually caused the club to fissure into two clubs.  The social interactions of the clubs' members outside of the official meetings were examined, and edges between members indicate that they interacted socially outside of the club setting.  The ground truth for this test is taken to be which specific club the members joined after the fissure.  

This example  illustrates a fundamental and intrinsic difficulty with the community detection problem: the definition of a community is problem dependent, and one can only design algorithms that are optimal for certain classes of communities. The communities on this network are defined in terms of who leads them, where the leaders can easily be identified by the two nodes with substantially higher degrees than the average of those they share connections with.  This suggests that a node's perspective on community should be defined by the leader(s) it is connected to, and the community scale features is defined by the perspective of its leader node.  If one were earnestly interested in solving community detection problems of this type, a very simple approach would be to take the edges involving the two nodes of highest degree as edge descriptor sets, agglomerate these based on which leader node is involved to form communities, and then assign any unclustered nodes to the community they have the most links to.  We have implemented this idea, and our algorithm yields a perfect recall value for each group, with F-scores both being over 0.94 for the given gold standard groupings.  

\begin{figure}[h!] 
 \begin{center}
	\includegraphics [width=0.50\textwidth] {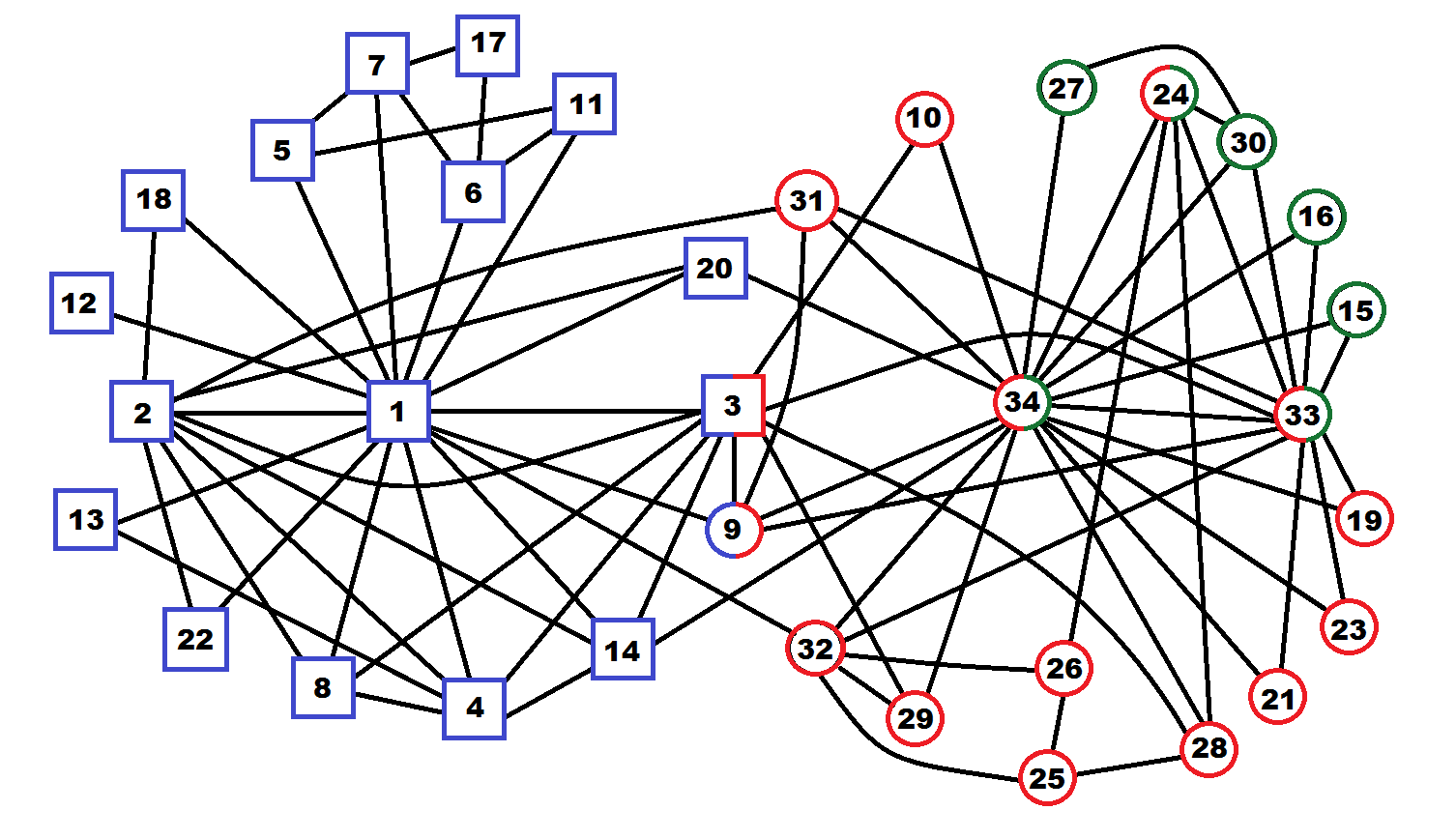} 
  \end{center}
   \caption{The Zachary karate club network.  The gold standard grouping for a node is given by its shape, and its found grouping is given by its color(s). }
\label{fig:OriginalKarateClubNetwork}
\end{figure}

Although this type of community structure is not at all what is intended for our algorithm to detect, it is a standard enough test to warrant seeing how it performs nonetheless.  In order to apply our algorithm to this network, we first need to get an initial estimate of what the community density should be.  To this end, we examine the edge density of the egonets for each node to get a local understanding of the average edge density of the network.  Finding that the average egonet link density is 78.2\%, we then set the community density to 3/4 of that in order to hold the communities to looser standards.  This results in the three clusters of nodes given below, with the precision, recall, and F-scores for these groups are presented in Table \ref{tab:Karate}.

%\pagebreak

\noindent Group 1: ~  1, 2, 3, 4, 5, 6, 7, 8, 9, 11, 12, 13,  \\
\indent \indent 14, 17, 18, 20, 22

\noindent Group 2: ~  3, 9, 10, 19, 21, 23, 24, 25, 26, 28, 29, \\
\indent \indent 31, 32, 33, 34

\noindent Group 3: ~ 15, 16, 24, 27, 30, 33, 34

%\noindent Group 1:  Precision =  0.94;  Recall =  1.0;  F = 0.97
%
%\noindent Group 2:  Precision =  0.93;  Recall =  0.78;  F = 0.85
%
%\noindent Group 3:  Precision =  1.0;  Recall =  0.39;  F = 0.56

\begin{table}[h]
\begin{center}
\begin{tabular}{|c|c|c|c|}
\hline
        & Precision & Recall & F-Score  \\ \hline
Group 1 & 0.94      & 1.0    & 0.97     \\ \hline
Group 2 & 0.93      & 0.78   & 0.85     \\ \hline
Group 3 & 1.0       & 0.39   & 0.56     \\ \hline
\end{tabular}
\caption{The precision, recall, and F-scores for the detected communities on the karate club network.  }
\label{tab:Karate}
\end{center}
\end{table}

As we can see, the communities produced by the algorithm cause the gold standard grouping, denoted by circles, to be split into two groups.  The reason for this splitting is that the network is mainly composed of two subtrees (one for each leader), and therefore the density of connection within each subtree remains low. Our approach, which assumes a more "egalitarian" community structure, is no longer optimal when the network is organized in such a strongly hierarchical way.   Splitting the gold standard group allows these new groups to have a higher edge densities of 34\% and 63\% whereas the community given as the gold standard only has an edge density of 27\%.  As we can see, although the division is not desired for this particular gold standard grouping, it is still a sensible one with respect to the notion of community our algorithm is designed to capture.

\subsection{High School Friendship Network}

\begin{figure}[h!] 
 \begin{center}
	\includegraphics [width=0.50\textwidth] {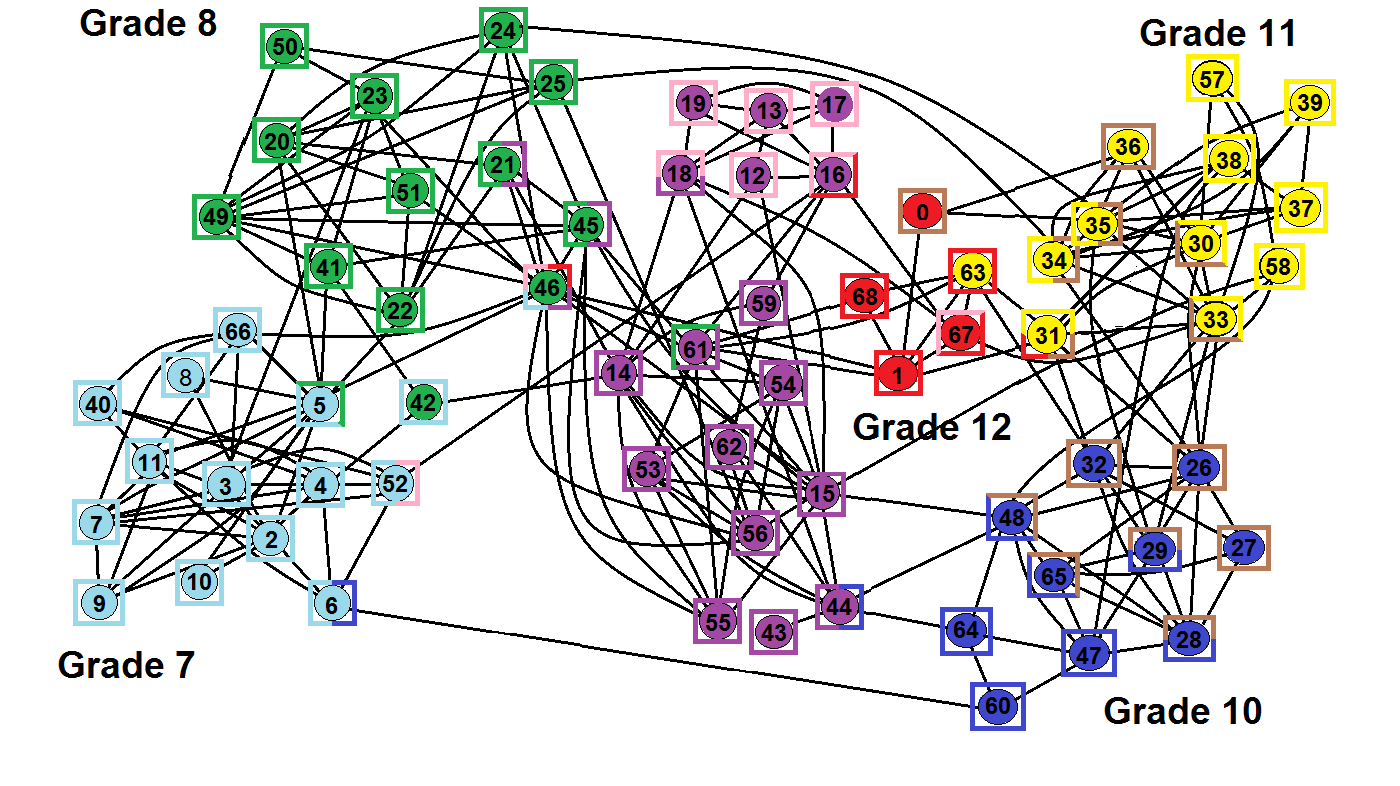} 
  \end{center}
   \caption{The high school friendhip network examined in Xie et al. 2013 \citep{Xi13}.  The ground truth for this network is reflected by the color coding of the nodes, and the found grouping for each node is reflected by the color(s) of the square surrounding it.}
\label{fig:HighSchoolFriendshipNetwork}
\end{figure}

We now describe the second real world network, used as a benchmark in a recent study that evaluated the states of the are algorithms for detecting overlapping communities \citep{Xi13}.  The dataset is part of the National Longitudinal Study of Adolescent to Adult Health \footnote{http://www.cpc.unc.edu/projects/addhealth/}.  The network is composed of high school students, where the links between students come from self-reported connections and the gold standard partitioning of the network is taken as the grades (7 through 12) the students belong to.  Although the ground truth is taken as six communities, it is understood that the friendship connections for grade 9 demonstrate that the grade can be split into two distinct subgroups with one group composed of black students and the other white students, as can be inferred from Figure \ref{fig:HighSchoolFriendshipNetwork}.

Our approach to this network is the same as the karate club network discussed in the previous section.  We estimate the desired community density by examining the average local edge density coming from each node's egonet, and set the community link density to 3/4 of that density.  For this network, the average egonet link density is found to be 67.0\%, so the community density threshold is set to 50.3\%.  We then take any nodes that remain unclustered after the community formation process, and assign them to the community they have the most links to.   The performance of our algorithm on this network is presented in Table \ref{tab:HighSchool}.

%\begin{table}[h!]
%\begin{center}
%\begin{tabular}{|c|c|c|c|c|c|}
%\hline
%Number of      & Number of               &                  &             &                  &          \\ 
% Communities & Overlapping Nodes  & Precision   & Recall   & F-Score    & NMI    \\ \hline
%          8                       &      20                                     &     0.79      & 0.82     & 0.80         &  0.52   \\ \hline
%\end{tabular}
%\caption{The performance of our algorithm on the high school friendship network.  }
%\label{tab:HighSchool}
%\end{center}
%\end{table}

\begin{table}[h]
\begin{center}
\begin{tabular}{ccc}
\hline
\multicolumn{1}{|c|}{\begin{tabular}[c]{@{}c@{}}Number of\\ Communities\end{tabular}} & \multicolumn{1}{c|}{Overlapping Nodes} & \multicolumn{1}{c|}{NMI}     \\ \hline
\multicolumn{1}{|c|}{8}                                                               & \multicolumn{1}{c|}{20}                & \multicolumn{1}{c|}{0.52}    \\ \hline
                                                                                      &                                        &                              \\ \hline
\multicolumn{1}{|c|}{Precision}                                                       & \multicolumn{1}{c|}{Recall}            & \multicolumn{1}{c|}{F-Score} \\ \hline
\multicolumn{1}{|c|}{0.79}                                                            & \multicolumn{1}{c|}{0.82}              & \multicolumn{1}{c|}{0.80}    \\ \hline

\end{tabular}
\caption{The performance of our algorithm on the high school friendship network.  }
\label{tab:HighSchool}
\end{center}
\end{table}

Despite its low NMI score, the groupings found by the algoritm are sensible ones.  It accurately detects the sub-division in grade 9, as well as the set of nodes falling in grades 9 and 10 that are densely connected to one another.  Additionally, if one examines the nodes which are "incorrectly" labelled (e.g. 0, 42, and 63), it is readily apparent that the gold standard groupings do not accurately represent the nature of how these nodes are connected to the network.

\section{Conclusion and Discussion of Results}

This work has focused on developing a computationally inexpensive algorithm capable of detecting overlapping communities in social networks.  A novel feature of how we approach the problem is that we define the community structure we are trying to capture based on how that structure would appear at differing scales.  The scales specifically considered in this paper are: the scale of individual nodes, the scale of individual communities, and the scale of the network as a whole.  Using the models developed in this work for each of these three scales, we find that applying our algorithm to benchmark tests demonstrates good overall performance.  This performance improves with increasing either the number of communities or the sizes of the communities to be detected.

One advantage of our methodology is that it explicitly accounts for multiscale features during the community formation process.  This aspect of our approach ensures that the detected communities are always sensible ones with respect to those features.  Another distinct advantage of our method is that the way we quantify the features at each scale and tie them together is highly modular.  This allows for the mathematical model of the community structure at any specific scale to be swapped out as appropriate based on the nature of a specific community detection problem.  

Future work will focus on further developing the methodology used in our algorithm.  One facet meriting further attention is to take advantage of the modularity of our algorithm to incorporate models of alternative features of community structure.  The potential advantage of this was demonstrated in Section \ref{sec:Karate}, where detecting the leader based communities of the karate club network became trivialized by modeling community features as appropriate to the problem.  Another avenue to explore is the possibility of chaining together sequences of node versus community level features, where community scale features are treated as node scale features at each higher link in the chain.  This will allow us to incorporate detection of hierarchical community structures into our algorithm, and further increase its flexibility.

\bibliographystyle{plainnat}

%\bibliographystyle{apalike}
%\bibliography{CommunityDetection_Biblio}

\end{document}